\documentclass[acmsmall,screen]{acmart}
\setkeys{acmart.cls}{urlbreakonhyphens=false}
\usepackage{tikz}\usetikzlibrary{positioning,arrows.meta,shapes.geometric}
\renewcommand\footnotetextcopyrightpermission[1]{}
\AtBeginDocument{}
\usepackage{booktabs}
\usepackage{array}
\usepackage{colortbl}
\newcommand{\nmodels}{16}

\newcommand{\nskills}{3,159}
\newcommand{\nrepos}{856}
\newcommand{\npairs}{2,608}

\newcommand{\naddedrepos}{200}
\newcommand{\ndeleted}{245}

\newcommand{\ndilution}{1,349}

\newcommand{\eqbound}{0.077}
\newcommand{\ninteractions}{12}
\newcommand{\maineffect}{0.41}
\newcommand{\eqfrac}{a fifth}
\newcommand{\hneedgain}{0.10}

\newcommand{\inplo}{18}
\newcommand{\inphi}{19}
\newcommand{\overheadpct}{20}
\newcommand{\helpedpct}{58}
\newcommand{\famlist}{Llama, Mistral, OLMo, Phi, Qwen}
\newcommand{\closedsonnetthreefivegain}{+0.42}

\newcommand{\closedhaikufourfivegain}{+0.45}

\newcommand{\closedopenlo}{+0.35}
\newcommand{\closedopenhi}{+0.46}

\newcommand{\closedeqbound}{0.062}

\newcommand{\agenticgainlo}{+0.18}
\newcommand{\agenticgainhi}{+0.33}

\newcommand{\intentn}{120}
\newcommand{\intentkappa}{0.60}

\newcommand{\intentkappalo}{0.43}

\newcommand{\intenthumanrate}{4.2}

\newcommand{\intentnew}{28}
\newcommand{\intentsync}{20}
\newcommand{\intentcorr}{18}

\newcommand{\intentmajk}{0.61}

\newcommand{\intentcommits}{6{,}957}
\newcommand{\intentstrict}{0.4}

\newcommand{\valn}{240}

\newcommand{\ddup}{\ensuremath{-}0.12}

\newcommand{\ddupn}{27}
\newcommand{\drep}{\ensuremath{-}0.58}

\newcommand{\drepn}{24}

\newcommand{\ctlnodirmeanabs}{0.21}

\newcommand{\xbntasks}{80}

\newcommand{\sotamold}{0.62}

\newcommand{\sotamnewold}{+0.19}

\newcommand{\sotaqnewold}{+0.16}

\newcommand{\latqapct}{\ensuremath{-}1.6}

\newcommand{\latpbpct}{\ensuremath{-}7.2}

\newcommand{\coevskillassetsmdiff}{+25}

\newcommand{\coevmedskill}{7}
\newcommand{\coevmedbase}{2}

\newcommand{\xbcn}{36}

\newcommand{\xbcmsfinold}{0.67}

\newcommand{\hvkflo}{0.50}
\newcommand{\hvkfhi}{0.55}

\newcommand{\hpk}{0.96}

\newcommand{\hthreek}{0.90}

\newcommand{\hspleast}{+0.09}

\newcommand{\hspmost}{+0.16}

\newcommand{\hsq}{+0.14}

\newcommand{\hsp}{+0.13}
\newcommand{\hsplo}{+0.05}
\newcommand{\hsphi}{+0.20}

\newcommand{\hson}{+0.06}
\newcommand{\hsonlo}{\ensuremath{-}0.06}
\newcommand{\hsonhi}{+0.18}

\newcommand{\needllmpairs}{1{,}454}

\newcommand{\abspathsd}{0.44}
\newcommand{\abscmdsd}{0.25}
\newcommand{\abswordsd}{101}
\newcommand{\uiaftern}{1{,}718}
\newcommand{\uiafterfirst}{3.3}
\newcommand{\uiafterlast}{6.7}
\newcommand{\uiallfirst}{2.2}
\newcommand{\uialllast}{5.9}
\newcommand{\revlateshare}{66}

\newcommand{\flagrepos}{146}
\newcommand{\flagpairs}{4{,}726}
\newcommand{\flagstars}{4{,}872}

\newcommand{\flagmodels}{16}
\newcommand{\flaggainlo}{+0.36}
\newcommand{\flaggainhi}{+0.54}

\newcommand{\discA}{+0.08}

\newcommand{\discFtasks}{36}
\newcommand{\discfull}{+0.20}

\newcommand{\famq}{+0.43}
\newcommand{\famo}{+0.39}
\newcommand{\famd}{+0.04}
\newcommand{\famdlo}{+0.02}
\newcommand{\famdhi}{+0.06}

\newcommand{\ovdirn}{55}
\newcommand{\ovred}{27}
\newcommand{\ovineff}{28}

\newcommand{\hneed}{25}
\newcommand{\hneedlo}{18}
\newcommand{\hneedhi}{46}

\newcommand{\mflo}{+0.02}
\newcommand{\mfhi}{+0.06}

\newcommand{\mfrello}{+0.38}
\newcommand{\mfrelhi}{+0.45}

\newcommand{\mfkept}{260}

\newcommand{\devA}{+0.36}

\newcommand{\devF}{+0.24}
\newcommand{\devFlo}{+0.07}
\newcommand{\devFhi}{+0.44}
\newcommand{\devdisc}{+0.13}

\newcommand{\msglo}{+0.03}
\newcommand{\msghi}{+0.08}

\newcommand{\texn}{154}

\newcommand{\texd}{+0.017}
\newcommand{\texdlo}{+0.002}
\newcommand{\texdhi}{+0.034}

\newcommand{\texmaxbound}{0.06}
\newcommand{\texpos}{+0.57}

\newcommand{\rcrulemed}{+248}

\newcommand{\rcprosemed}{+6}

\newcommand{\texin}{+34}

\newcommand{\texout}{+2}

\newcommand{\unchn}{300}

\newcommand{\phixedge}{+0.035}

\newcommand{\phixedgemain}{+0.028}
\newcommand{\phixn}{150}
\newcommand{\phixnon}{10}
\newcommand{\phixlo}{+0.26}
\newcommand{\phixhi}{+0.42}

\newcommand{\phixdd}{+0.007}
\newcommand{\phixddlo}{\ensuremath{-}0.029}
\newcommand{\phixddhi}{+0.042}

\newcommand{\presold}{0.61}
\newcommand{\newtokold}{0.05}

\newcommand{\retnegin}{22}
\newcommand{\retnegout}{32}
\newcommand{\retcondin}{25}
\newcommand{\retcondout}{32}

\newcommand{\pdnetin}{+123}
\newcommand{\pdnetall}{+18}
\newcommand{\pdcomin}{4}
\newcommand{\pdcomall}{2}

\newcommand{\needcilo}{14}
\newcommand{\needcihi}{40}
\newcommand{\needgainlo}{0.05}
\newcommand{\needgainhi}{0.17}
\newcommand{\discutilq}{61}
\newcommand{\discutilqlo}{48}
\newcommand{\discutilqhi}{74}

\newcommand{\bodyutilq}{62}
\newcommand{\discvisq}{71}
\newcommand{\discreadq}{89}
\newcommand{\discutils}{88}
\newcommand{\discutilslo}{77}
\newcommand{\discutilshi}{97}

\newcommand{\bodyutils}{81}
\newcommand{\discviss}{69}
\newcommand{\discreads}{88}
\newcommand{\discutilgap}{+0.24}
\newcommand{\discutilgaplo}{+0.12}
\newcommand{\discutilgaphi}{+0.36}

\newcommand{\discretlo}{16}
\newcommand{\discrethi}{61}

\newcommand{\lcretseslo}{53}
\newcommand{\lcretseshi}{72}

\newcommand{\lcretlonglo}{29}
\newcommand{\lcretlonghi}{46}

\newcommand{\xbpresq}{+0.00}

\newcommand{\xbpresqq}{+0.06}

\newcommand{\xbpresm}{\ensuremath{-}0.13}

\newcommand{\xbpresn}{23}

\newcommand{\supmatrix}{S1}
\newcommand{\supvalidation}{S2}
\newcommand{\supannot}{S3}
\newcommand{\suprepr}{S4}
\newcommand{\supretained}{S5}
\newcommand{\supprobedist}{S6}
\newcommand{\suprevisions}{S7}
\newcommand{\supgrowth}{S8}
\newcommand{\supcoevo}{S9}
\newcommand{\supextractor}{S10}
\newcommand{\supnovelty}{S11}
\newcommand{\suptexture}{S12}
\newcommand{\suptypology}{S13}

\newcommand{\supdel}{S15}
\newcommand{\supdelcat}{S16}
\newcommand{\supequiv}{S17}
\newcommand{\supequivsens}{S18}
\newcommand{\supocc}{S19}
\newcommand{\supdilution}{S20}

\newcommand{\suprevcost}{S23}
\newcommand{\suplatency}{S24}
\newcommand{\supclosed}{S25}
\newcommand{\supclosedladder}{S26}
\newcommand{\supagentic}{S27}
\newcommand{\supxbenchcommon}{S28}
\newcommand{\supsota}{S29}
\newcommand{\supbehaviour}{S30}
\newcommand{\supdollars}{S31}
\newcommand{\supmisfire}{S32}
\newcommand{\supflagship}{S33}
\newcommand{\supgenerator}{S34}
\newcommand{\supneed}{S35}
\newcommand{\supneedci}{S36}

\newcommand{\supharnessfiles}{S39}
\newcommand{\supfiggrowth}{S1}

\newcommand{\hwordsup}{60}

\newcommand{\reprsnapmed}{0}
\newcommand{\reprhistmed}{30}

\newcommand{\nvrest}{18}
\newcommand{\nvexn}{7}

\newcommand{\nvex}{+0.04}
\newcommand{\nvexlo}{\ensuremath{-}0.01}
\newcommand{\nvexhi}{+0.08}

\newcommand{\rdnotold}{57}
\newcommand{\rdseenlo}{42}
\newcommand{\rdseenhi}{50}

\newcommand{\rdsawqq}{+0.19}
\newcommand{\rdsawqqlo}{+0.08}
\newcommand{\rdsawqqhi}{+0.29}
\newcommand{\agcostlo}{\ensuremath{-}2.3}
\newcommand{\agcosthi}{+3.0}
\newcommand{\agcostcilo}{\ensuremath{-}14}
\newcommand{\agcostcihi}{+12}

\newcommand{\lcsig}{8}
\newcommand{\lcrows}{8}
\newcommand{\lcseslen}{15}
\newcommand{\lclonglen}{40}

\newcommand{\lcprobes}{149}
\newcommand{\lcstrange}{+0.37 to +0.42}
\newcommand{\lcgainrange}{+0.07 to +0.28}
\newcommand{\ndiscopen}{three}

\newcommand{\discpoolret}{38}
\newcommand{\discnineA}{+0.07}

\newcommand{\discson}{+0.19}

\newcommand{\discsonF}{+0.19}
\newcommand{\discsonFlo}{+0.03}
\newcommand{\discsonFhi}{+0.39}
\newcommand{\discsonFold}{0.50}
\newcommand{\discsonFnew}{0.69}

\newcommand{\flatpct}{17}
\newcommand{\downpct}{5}
\newcommand{\nmain}{300}
\newcommand{\nmainorig}{245}
\newcommand{\nmaindrawn}{55}
\newcommand{\nextfit}{59}

\newcommand{\ncandall}{2{,}300}
\newcommand{\npool}{396}
\newcommand{\nkeeppct}{17}
\newcommand{\truncn}{12{,}672}
\newcommand{\truncover}{618}

\newcommand{\occlow}{0.424}
\newcommand{\occhigh}{0.270}
\newcommand{\novnewlo}{+0.51}
\newcommand{\novnewhi}{+0.66}

\newcommand{\ovcilo}{16.1}
\newcommand{\ovcihi}{24.2}

\newcommand{\ovthrlo}{18}
\newcommand{\ovthrhi}{27}

\newcommand{\famon}{10}
\newcommand{\corpusmaxstars}{375}
\newcommand{\fnlines}{143{,}196}
\newcommand{\fnimp}{12{,}564}
\newcommand{\fntok}{4{,}041}
\newcommand{\apresn}{107}

\newcommand{\apresd}{+0.08}
\newcommand{\apreslo}{+0.03}
\newcommand{\apreshi}{+0.12}

\newcommand{\rvt}{+0.15}
\newcommand{\rvtlo}{+0.12}
\newcommand{\rvthi}{+0.19}
\newcommand{\rvtapres}{+0.04}
\newcommand{\rvtapreslo}{+0.02}
\newcommand{\rvtapreshi}{+0.07}
\newcommand{\rvtapresn}{88}

\newcommand{\rvtbound}{0.071}

\newcommand{\ntstep}{Qwen 2.5-7B to 3-8B}
\newcommand{\ntint}{\ensuremath{-}0.083}

\newcommand{\ntholm}{0.018}

\newcommand{\densmain}{+0.43}
\newcommand{\densmainlo}{\ensuremath{-}0.10}
\newcommand{\densmainhi}{+0.89}
\newcommand{\densflag}{+0.08}
\newcommand{\densflaglo}{\ensuremath{-}0.21}
\newcommand{\densflaghi}{+0.66}
\newcommand{\linkminholm}{0.24}
\newcommand{\mainthreeoh}{0.405}
\newcommand{\mainthreeofour}{0.401}
\newcommand{\dfkappa}{0.86}
\newcommand{\tckappa}{0.72}
\newcommand{\dfeffa}{+0.52}

\newcommand{\dfpreclo}{0.97}
\newcommand{\dfprechi}{0.98}

\newcommand{\dfrecaone}{0.73}
\newcommand{\dfrecatwo}{0.82}
\newcommand{\dfrecold}{0.79}
\newcommand{\dfrecnew}{0.65}

\newcommand{\dfmech}{+0.41}
\newcommand{\tcdefn}{163}
\newcommand{\tcconf}{23}
\newcommand{\tcunc}{77}
\newcommand{\tcadjn}{100}
\newcommand{\dfdefn}{197}
\newcommand{\sonpower}{0.5}
\newcommand{\sonmde}{0.18}
\newcommand{\sonpairs}{72}
\newcommand{\sonpairsneed}{125}
\newcommand{\lcin}{+0.31}
\newcommand{\lcinlo}{+0.20}
\newcommand{\lcinhi}{+0.42}
\newcommand{\lcinn}{28}

\newcommand{\authshare}{37.6}

\newcommand{\authnonskill}{40.5}

\newcommand{\mtpairs}{3.3}
\newcommand{\clfkappa}{0.70}
\newcommand{\clfacc}{0.76}
\newcommand{\clfcodeprec}{0.55}
\newcommand{\clfmustrec}{0.18}
\newcommand{\clfcondrec}{0.19}

\newcommand{\expathprec}{1.00}
\newcommand{\expathrec}{0.21}
\newcommand{\excmdprec}{0.93}
\newcommand{\excmdrec}{0.13}
\newcommand{\extoolprec}{0.15}

\newcommand{\revrulepct}{55}
\newcommand{\revrulepctx}{55.0}

\newcommand{\revcodepct}{19}
\newcommand{\revprosepct}{14}
\newcommand{\revnpairs}{2,608}
\newcommand{\hfrevn}{6{,}955}
\newcommand{\hfrevrate}{29}
\newcommand{\hfbaserate}{10}
\newcommand{\hfmrate}{23.0}
\newcommand{\hfmbase}{18.5}
\newcommand{\hfmdiff}{+4.5}
\newcommand{\hfmlo}{+1.6}
\newcommand{\hfmhi}{+7.3}
\newcommand{\hfcrmrate}{36.7}
\newcommand{\hfcrmbase}{21.8}
\newcommand{\hfcrmdiff}{+14.8}
\newcommand{\hfcrmlo}{+9.4}
\newcommand{\hfcrmhi}{+20.4}

\newcommand{\hfnrepos}{855}

\newcommand{\hfcmddiff}{+2.1}

\newcommand{\hfamddiff}{+2.0}

\newcommand{\hfcmdsdiff}{+1.4}

\newcommand{\hfagtdiff}{+1.4}

\newcommand{\hfothdiff}{+1.9}

\newcommand{\stall}{+0.23}

\newcommand{\stres}{+0.22}
\newcommand{\streslo}{+0.14}
\newcommand{\streshi}{+0.30}
\newcommand{\stresn}{62}

\newcommand{\ststalen}{18}
\newcommand{\stdiff}{\ensuremath{-}0.05}
\newcommand{\stdifflo}{\ensuremath{-}0.24}
\newcommand{\stdiffhi}{+0.16}

\newcommand{\stresmin}{+0.13}
\newcommand{\stresmax}{+0.35}
\newcommand{\stallgain}{0.23}
\newcommand{\nsamprepos}{1{,}500}
\newcommand{\nsampfiles}{7{,}959}
\newcommand{\nsampshare}{40}
\newcommand{\faact}{+0.23}
\newcommand{\faactlo}{+0.15}
\newcommand{\faacthi}{+0.31}
\newcommand{\faod}{+0.12}
\newcommand{\faodlo}{+0.06}
\newcommand{\faodhi}{+0.18}
\newcommand{\fahum}{+0.10}
\newcommand{\fahumlo}{+0.04}
\newcommand{\fahumhi}{+0.17}
\newcommand{\faret}{51}
\newcommand{\faretlo}{28}
\newcommand{\farethi}{75}

\newcommand{\facarry}{50}

\newcommand{\famarg}{+0.8}

\newcommand{\stneedgain}{0.06}
\newcommand{\agskillloabs}{37}
\newcommand{\agskillhiabs}{57}
\newcommand{\dtokopenabs}{17}
\newcommand{\dtokclosedloabs}{24}
\newcommand{\dtokclosedhiabs}{27}
\newcommand{\unchlineabs}{0.24}
\newcommand{\insight}[1]{\par\vskip\smallskipamount\noindent\fcolorbox{black!45}{black!4}{\parbox{\dimexpr\linewidth-2\fboxsep-2\fboxrule\relax}{\textbf{Insight.} #1}}\par\vskip\smallskipamount}

\begin{document}

\title[Agent Skill Evolution]{Agent Skill Evolution: How Revisions Affect Coding Agents}

\author{Jiajie Wang}
\affiliation{
  \institution{The University of Arizona}
  \country{USA}
}

\author{Yutong Zhao}
\affiliation{
  \institution{California State University, Long Beach}
  \country{USA}
}

\author{Tianlin Li}
\affiliation{
  \institution{Nanyang Technological University}
  \country{Singapore}
}

\author{Huashan Chen}
\affiliation{
  \institution{Institute of Information Engineering, Chinese Academy of Science}
  \country{China}
}

\author{Jinfu Chen}
\affiliation{
  \institution{Wuhan University}
  \country{China}
}

\author{Kebin Peng}
\affiliation{
  \institution{East Carolina University}
  \country{USA}
}

\author{Sen He}
\affiliation{
  \institution{The University of Arizona}
  \country{USA}
}

\begin{abstract}
Agent Skills, the \texttt{SKILL.md} files that tell an LLM coding agent how a project works, are revised like code, yet what a revision
does to the agent is unknown. From \npairs{} first/last revision pairs of \nskills{} Skills, we characterize how Skills evolve and how they
change together with the configuration of the agent's harness. We then focus on \emph{rule changes}, revisions that add or remove a rule we
can check automatically, such as ``run \texttt{allium check}''. We measure their effect on 21 models in single answers and on four agents in
a sandbox, and their cost on 20 of these models and the four agents. Most revisions (\revrulepct\%) change a rule or procedure, and commits that revise a Skill change harness files such as
\texttt{CLAUDE.md} more often than other commits of the same size. Across 16 open-weight models, an added rule raises compliance in a single
answer by +\maineffect{} on average. Across the four agents, the rate at which the agent takes the required action rises by \faact{} on
average (\sotaqnewold{} to \devA), and for the three agents that blind judges assessed, final correctness rises by \fahum{} on average
(\hson{} to \hsq). The gain comes mainly from rules
that name a command or path the old Skill did not mention. Real tools load a Skill's body only when the agent decides it needs it. In
that setting the four agents keep about half of the action gain on average (\faret\%), and the \ndiscopen{} open models about
\discpoolret\%. A revision adds \inplo--\inphi\% input tokens to a single answer and no detectable cost to an agent episode, while loading
a Skill's body raises the tokens of an episode by \facarry\% on average.
\end{abstract}

\maketitle

\section{Introduction}

In the agent literature, a \emph{skill} is a reusable procedure that an agent retrieves and applies, stored as code, as natural-language
instructions, or as a packaged directory~\cite{wang2023voyager,li2026dynamic}. An \emph{Agent Skill} is the packaged form introduced by
Anthropic and released as an open standard: a folder whose \texttt{SKILL.md} file starts with a name and a description in YAML
\emph{frontmatter}, a short header block, followed by Markdown instructions and optionally by scripts and reference
files~\cite{anthropic2025skills,agentskills2026spec}. A Skill reaches the model through a \emph{harness}, the software layer around a
language model that supplies its tools, manages the context the model sees, and runs the loop that chains its actions. Model and harness
together form the agent~\cite{anthropic2026ccglossary,anthropic2026evals}. The repository files that configure a harness, such as
\texttt{CLAUDE.md}, \texttt{AGENTS.md}, settings, hooks and MCP configuration, we call \emph{harness configuration}. Harnesses load Skills
in stages, by \emph{progressive disclosure}. Each Skill's name and description sit in the system prompt, the body enters the context when
the Skill is activated, and bundled files are read or run only when needed~\cite{anthropic2026skillsdocs,agentskills2026client}. Public
registries list well over a hundred thousand Skills, and evaluations disagree. Curated Skills raise pass rates on one
benchmark~\cite{li2026skillsbench} but barely on another~\cite{han2026sweskills}, Skills distilled from agent trajectories add only a few
points over no Skill~\cite{jiang2026demystifying}, and the token overhead can be severe~\cite{dong2026harmful}.

Most of that work evaluates a Skill at one version~\cite{li2026skillsbench,han2026sweskills,liu2026wild,zhou2026counterfactual}. Yet Skills are files in a repository, and maintainers revise them like other files.
They change commands, delete stale sections and add checklists. Among the Skills we study, the first and last observed versions are a
median of two revision commits apart, and \hwordsup\% of revised Skills end longer than they started. Every edit changes what the agent is
told and how many tokens it reads. \textbf{Yet the evolution of Agent Skills and its effects on agents remain understudied.}

To address this gap, we conduct, to the best of our knowledge, the first large-scale empirical study of Agent Skill evolution and its
effects on agents. We study the rules that maintainers added or removed in real revisions, and we compare the old and new Skill on the same
request and model. We use \emph{revision} for the cumulative change between the first and the
last observed version of a Skill, which may span several \emph{revision commits}. We study \emph{rule changes}, revisions that add or
delete a concrete, mechanically checkable rule that names a command, path or identifier. We call the line that states such a rule a
\emph{directive}. We aim to exclude rewording of an existing rule; line-level extraction still admits some restated rules, which
\S\ref{sec:novelty} measures separately. Models run without thinking modes, and single-step prompts are capped
at 8{,}192 tokens of usable input (\S\ref{sec:settings}, \S\ref{sec:metrics}). Each probe is a request built so that the rule applies, and
we measure what a revision does when its rule is needed. We ask three research questions. RQ1 comes first because it shows that most revisions
change a rule or procedure and that Skills change together with harness configuration. RQ2 and RQ3 then measure what a narrower class of
rule changes does, rules that name a command, path or identifier.

\textbf{RQ1: How do Agent Skills co-evolve with their harness?} Revised Skills change often, and most revisions
change a rule or procedure. Executable references increase in a small fraction of Skills. Matched by repository and commit-size band,
commits that revise a Skill change harness configuration more often than other commits, chiefly \texttt{CLAUDE.md}, \texttt{AGENTS.md} and
agent and command definitions, and commits that create a Skill do so more often still. Few revisions document a model-related motive. Over
a third of revision commits carry an AI-agent or bot signature, a share close to that of other commits.

\textbf{RQ2: How does Skill evolution affect task effectiveness?} An added rule raises compliance in single answers on all 21 models, and
stating the rule adds compliance beyond naming its token. The gain comes mainly from rules that name a command or path the old Skill did
not mention. Seeing that token elsewhere in the context or the repository does not replace the rule. In a sandbox, agents take
the required action more often, even when every file the old Skill names still exists, and blind judges see modestly higher final
correctness. Older and newer model generations gain similarly. Effects are smaller in
the agent settings we test. With on-demand loading the four agents keep about \faret\% of the action gain on average. Claude Sonnet~4.5's
gain is \discson{} under both loading modes, and \ndiscopen{} open models keep about \discpoolret\%.

\textbf{RQ3: How does Skill evolution affect cost?} A revision lengthens the prompt but shortens the answer. Inside an agent episode we
could not detect extra cost from the revision, while loading the Skill body at all adds \agskillloabs{} to \agskillhiabs\% more tokens. One
rule--model pair in five adds tokens without a detectable gain.

\paragraph{This work makes the following contributions:}
\begin{itemize}
\item We conduct a large-scale empirical study of Agent Skill evolution: we reconstruct the revision histories of \nskills{} Skills in
\nrepos{} repositories and characterize how their content, executable references and harness configuration change, and who writes the
revisions.
\item We measure the effects of Skill evolution on effectiveness on 21 models (\nmodels{} open-weight and five closed) and on cost on 20 of
them, plus an open coding-agent model, from single-step answers to multi-step agents in a sandbox, with control experiments and blind human judgement of
outcomes.
\item We make available a dataset of \nskills{} Agent Skills with their revision histories, \npool{} probes derived from real revisions, an
executable benchmark of \xbntasks{} tasks and the human judgements, together with a replication package (Data Availability).
\end{itemize}

\section{Related Work}
\subsection{Studies on Agent Skills}
Several recent studies examine Skills directly. Dong et al.~\cite{dong2026harmful} attribute agent failures to specific loaded Skills
with a differential design and trace most context-overhead failures to mandatory Skill-body text. Jiang et al.~\cite{jiang2026demystifying}
find that a distilled \texttt{SKILL.md} beats workflow memory built from the same trajectories by six points and that Skills act mostly as
\emph{procedural anchors} (65.7\% of cases) rather than as \emph{knowledge injection} (4.5\%). Wang et al.~\cite{wang2026implicit} show that coordinated wording in a
prompt and a Skill description can steer which Skill is selected, and Zhou et al.~\cite{zhou2026counterfactual} pair with- and without-Skill
traces on 49 SWE-Skills-Bench tasks and find that Skills reshape behaviour while the pass rate barely moves. Yang and Ding~\cite{yang2026webdevskills}
place \texttt{SKILL.md} in the prompt as the controlled variable with an irrelevant-Skill control of approximately equal byte length ($\pm$5\%), and Gao et al.~\cite{gao2026skillreducer}
find that over 60\% of Skill body text is not actionable. Zhang et al.~\cite{zhang2026skillmd} find reuse defects in most of 138{,}133
public \texttt{SKILL.md} files. These studies compare a Skill with its absence, or analyse it, at one version. Work on skill libraries
studies how an agent adds and refines its own skills~\cite{wang2023voyager,li2026dynamic}.

\subsection{Studies on Agent Harnesses}
Chatlatanagulchai et al.~\cite{chatlatanagulchai2025agentreadmes} studied 2{,}303 agent context files in 1{,}925 repositories and found that
they evolve like configuration code, through frequent small additions. Mohsenimofidi et al.~\cite{mohsenimofidi2026context} and Jiang and
Nam~\cite{jiang2025cursorrules} characterized how \texttt{AGENTS.md} files and Cursor rules present project context, and Galster
et al.~\cite{galster2026configuring} found that context files dominate agent configuration in 2{,}853 repositories while Skills remain rare
and mostly hold static instructions rather than executable scripts. Hereiz et al.~\cite{hereiz2026plugins} found that instruction files and scripts inside plugin Skills co-evolve, and prompts embedded in
code change primarily through additions and modifications~\cite{tafreshipour2025prompting}. Earlier, infrastructure-as-code files were found to co-evolve with the
source code they configure~\cite{jiang2015coevolution} and to exhibit configuration smells~\cite{sharma2016smells}, and we ask the same
question of Skills and harness configuration. On effect,
Gloaguen et al.~\cite{gloaguen2026agentsmd} found that context files do not generally raise task success but add over 20\% inference cost,
and McMillan~\cite{mcmillan2026adherence} detected no effect of file structure on adherence. Ben Sghaier et al.~\cite{bensghaier2026harness} held the model fixed across 35 harness
releases and found that tokens per task rose by over 70\% with no significant gain in resolve rate. Most prior work compares a configuration
file with its absence or treats the harness as one unit; we measure what one revision of a Skill changes in an agent's effectiveness and
cost.

\subsection{Studies on Agentic Cost}
Kapoor et al.~\cite{kapoor2025agents} argued that agent evaluations must be cost-controlled, and Erol et al.~\cite{erol2026costofpass}
formalized cost-of-pass, the expected cost of one correct solution. Input tokens drive the cost of SWE-bench trajectories, which vary up to 30-fold
on the same task~\cite{bai2026spend} and snowball over turns~\cite{fan2025sweeffi}, and several studies cut cost by shortening
trajectories~\cite{lindenbauer2025complexity,xiao2026agentdiet,gao2026morewithless}.

\section{Study Design}\label{sec:design}
\begin{figure*}[t]\centering
\includegraphics[width=\textwidth]{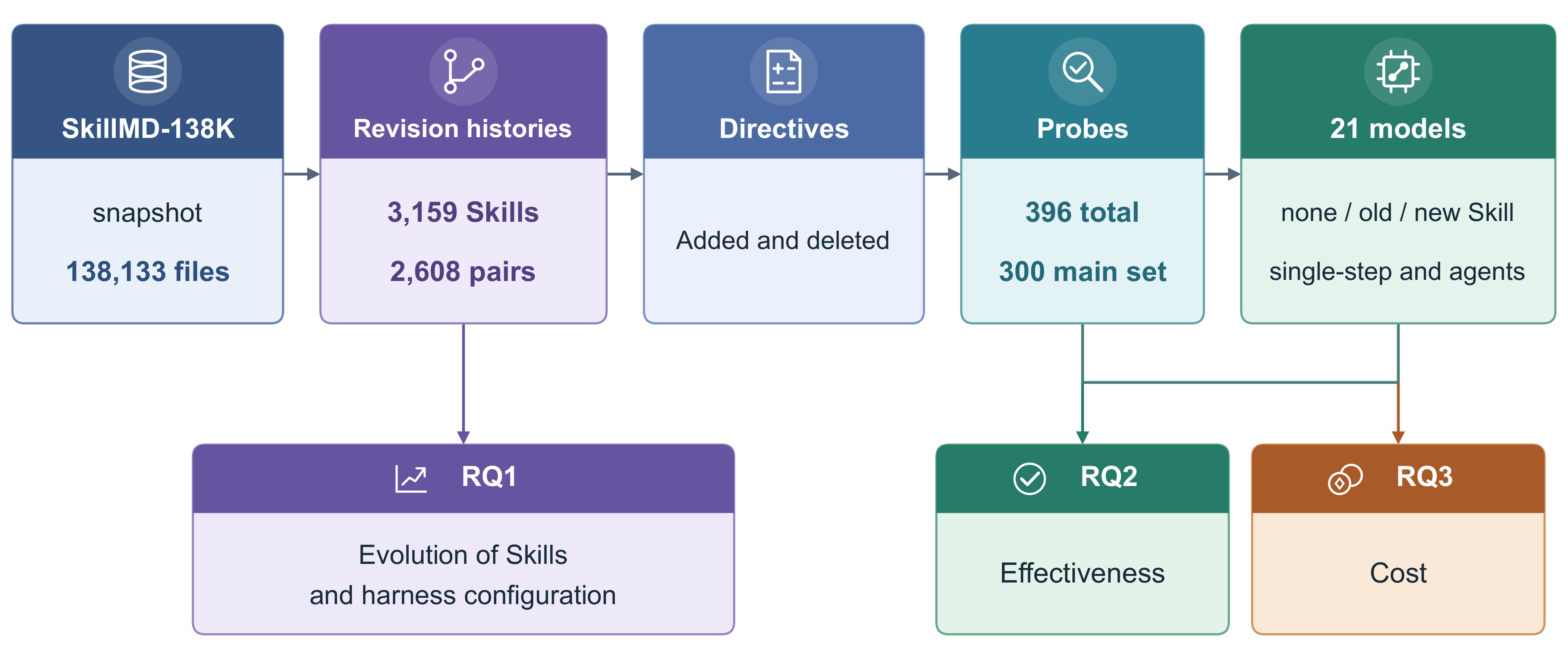}
\Description{Flow diagram of the study. The SkillMD-138K snapshot leads to revision histories, then to added and deleted directives,
then to probes, which are run on the models under the none, old and new conditions. Revision histories feed RQ1; probes and model runs feed
RQ2 on effectiveness and RQ3 on cost.}
\caption{Study design.}
\label{fig:overview}\end{figure*}

Figure~\ref{fig:overview} shows the pipeline, and Figure~\ref{fig:probe} follows one revision through it. We rebuild the revision history
of each sampled \texttt{SKILL.md} file and pair its first and last versions (\S\ref{sec:corpus}). RQ1 characterizes these pairs, the
commits behind them and the harness files changed with them. From each pair we extract the directives a maintainer added or deleted and
turn each into a \emph{probe}, a user request that needs the directive, with a string check for it (\S\ref{sec:probes}). We present every
probe with the Skill's name and description alone and with the old or the new body, to 21 models in single answers and to agents that act
over many steps, among them four agents in a sandbox (\S\ref{sec:settings}). We then score effectiveness at three levels and count tokens
and time (\S\ref{sec:metrics}). RQ2 compares effectiveness and RQ3 cost between the old and the new version.

\subsection{Dataset Construction}\label{sec:corpus}
We start from SkillMD-138K~\cite{skillmd138k,zhang2026skillmd}, a public snapshot of 138{,}133 \texttt{SKILL.md} files from 20{,}556
repositories, found through a Skill registry (by our count, 90\% of files), GitHub code search~\cite{github} (5\%) and direct clones (4\%; 1\% lack a
recorded channel). The snapshot has no history, so we rebuild it. We draw \nsamprepos{} repositories in star-stratified batches
(\nsampfiles{} \texttt{SKILL.md} files, at most 40 per repository), make a blobless clone of each within a fixed time budget, and follow
every \texttt{SKILL.md} through \texttt{git log -{}-follow}, so renames and moves are tracked. A \emph{version} is the file's content at one
commit that touches it, so a Skill with $v$ versions has $v-1$ revision commits after the commit that creates it. We keep files with at
least two versions, \nsampshare\% of the sampled files. Files in repositories we could not clone within the budget count as unrevised, so
this share is a lower bound.
For the version-pair analyses we keep those whose first and last texts differ (Table~\ref{tab:dataset}). A first/last pair spans a median
of three versions, that is, two revision commits (five versions behind our probes).

\begin{table}[t]\centering\caption{Dataset construction at a glance (counts in parentheses: repositories unless stated), with every hand-labelled sample used in the paper.}\label{tab:dataset}\footnotesize
\begin{tabular}{@{}lr@{}}\toprule
Stage & Count\\\midrule
SkillMD-138K snapshot: \texttt{SKILL.md} files & 138{,}133 (20{,}556)\\
Sampled repositories: \texttt{SKILL.md} files (at most 40 each) & \nsampfiles{} (\nsamprepos)\\
Skills with at least two versions & \nskills{} (\nrepos)\\
\quad first/last pairs whose text differs & \npairs\\
\quad revision commits & \intentcommits\\
High-star set: first/last pairs (no 40-file cap) & \flagpairs{} (\flagrepos)\\\midrule
Lines added across the pairs & \fnlines\\
\quad imperatives / naming a backticked token & \fnimp{} / \fntok\\
\quad directive candidates (at most three per Skill) & \ncandall\\
\quad added-directive probes passing the filter & \npool{} (\naddedrepos)\\
\quad main set, all 16 open-weight models & \nmain\\
Deleted-directive / dilution probes & \ndeleted{} / \ndilution\\
Executable sandbox tasks & \xbntasks{} (57)\\\midrule
Human labels: responses (two annotators, adjudication) & \valn\\
\quad commits (intent) / deleted directives (reason) & \intentn{} / 64\\
\quad later commit--rule pairs (need) / added lines (classifier / extractor) & 100 / 200 / 250\\
\quad sandbox artifacts (two annotators per round, adjudication) & 144 + 336\\
\bottomrule\end{tabular}\end{table}

\paragraph{Skill selection criteria}
Repositories whose Skills have a revision history are more popular than the snapshot as a whole (Table~\suprepr; median \reprhistmed{}
stars against \reprsnapmed). No repository in the main set has more than \corpusmaxstars{} stars, because larger clones exceeded our time
budget, so we cloned a separate \emph{high-star set} with a longer budget. It holds \flagrepos{} repositories above 1{,}000 stars (median
\flagstars), sampled in three star bands from the same snapshot, on which we repeat the RQ1 description and the central RQ2 comparison.

\paragraph{Commit labels}
Three LLMs of different families or sizes (Qwen~3.5-4B, phi-4, Qwen~3.5-27B) label each of the \intentcommits{} revision commits with
one of nine intent categories, and we keep the majority label. An annotator labelled a sample of \intentn{} commits by hand. The majority
label agrees with the annotator at $\kappa=\intentmajk$ (single LLMs \intentkappalo--\intentkappa), so we report shares from the annotator's
labels. Agent
authorship is read from commit metadata, namely co-author trailers, tool footers and bot accounts.

\subsection{Probe Construction}\label{sec:probes}
\paragraph{From a revision to a probe}
Figure~\ref{fig:probe} follows one revision through four steps.
\emph{(1)~Directive.} Our unit of analysis is a directive, a line a maintainer added or deleted that expresses an imperative and names a
concrete token (a command, flag, path or identifier) in backticks. A regular-expression pass selects such lines, at most three per Skill
(Table~\ref{tab:dataset}). In the figure, line~21 of the new version adds ``\textbf{Must read}: \texttt{knowledge/projects/\_index.md}''.
\emph{(2)~Request and check.} An LLM (Qwen~3.5-27B, greedy decoding) writes a realistic user request whose correct handling needs the
directive without mentioning it, and copies from the directive the strings that a compliant answer must contain (here the index path) or
must not contain. A filter keeps a probe if these strings occur verbatim in the directive, are absent from the request, are not generic
and do not contradict each other.
\emph{(3)~Conditions.} The probe is presented under \textbf{none}, \textbf{old} and \textbf{new} (\S\ref{sec:metrics}). The
\emph{estimand}, the quantity we estimate, is the new-minus-old difference on the same probe.
\emph{(4)~Grading.} A response complies if its code spans (the whole response if it has none) contain every required string,
case-insensitively, and no forbidden string outside a negated context, a verifiable-instruction test in the sense of IFEval~\cite{zhou2023ifeval}.
In the example, no model complies under none or with the old version, and 43 of 80 samples comply with the new one.

\begin{figure}[t]
\definecolor{fraddBg}{HTML}{CCDDAA}\definecolor{fraddFg}{HTML}{225522}%
\definecolor{frdelBg}{HTML}{FFCCCC}\definecolor{frdelFg}{HTML}{663333}%
\definecolor{frmodBg}{HTML}{EEEEBB}\definecolor{frmodFg}{HTML}{666633}%
\definecolor{frExeBg}{HTML}{CCEEFF}\definecolor{frExeFg}{HTML}{4477AA}%
\definecolor{frNum}{HTML}{777777}%
\newlength{\frch}\newlength{\frgut}\newlength{\frside}\newlength{\frmid}%
\newcommand{\frtt}{\fontsize{7}{8}\selectfont\ttfamily}%
\settowidth{\frch}{\frtt x}\setlength{\frgut}{13pt}\setlength{\frmid}{4pt}%
\setlength{\frside}{\dimexpr0.5\columnwidth-0.5\frmid-0.2pt\relax}%
\newcommand{\frsym}[1]{\csname frsym@#1\endcsname}%
\expandafter\def\csname frsym@add\endcsname{\textcolor{fraddFg}{\textbf{+}}}%
\expandafter\def\csname frsym@del\endcsname{\textcolor{frdelFg}{\textbf{-}}}%
\expandafter\def\csname frsym@mod\endcsname{\textcolor{frmodFg}{\textbf{\textasciitilde}}}%
\newcommand{\frx}[1]{\smash{\setlength{\fboxsep}{0.5pt}\setlength{\fboxrule}{0.5pt}\fcolorbox{frExeFg}{frExeBg}{\vphantom{Ag}#1}}}%
\newcommand{\frel}{\textcolor{frNum}{[...]}}%
\newcommand{\frprobe}{\hfill{\sffamily\fontsize{5.5}{6}\selectfont\textcolor{fraddFg}{$\leftarrow$\,probe}}}%
\newcommand{\frl}[4]{\par\nointerlineskip\noindent\colorbox{fr#1Bg}{\parbox[t]{\linewidth}{\frtt\raggedright
  \hyphenpenalty=10000\exhyphenpenalty=10000\hangindent=\dimexpr\frgut+#3\frch+2\frch\relax\hangafter=1
  \strut\makebox[\frgut][l]{\makebox[7pt][r]{\sffamily\fontsize{5}{6}\selectfont\textcolor{frNum}{#2}}\hspace{1.5pt}\frsym{#1}}%
  \hspace*{#3\frch}#4\strut}}}%
\newcommand{\frhunk}[2]{\par\noindent\mbox{\begin{minipage}[t]{\frside}\setlength{\fboxsep}{0pt}#1\end{minipage}%
  \hspace{\frmid}\begin{minipage}[t]{\frside}\setlength{\fboxsep}{0pt}#2\end{minipage}}\par\vspace{0.7pt}}%
\newcommand{\frhead}[2]{\par\noindent\mbox{\begin{minipage}[t]{\frside}\sffamily\scriptsize #1\end{minipage}%
  \hspace{\frmid}\begin{minipage}[t]{\frside}\sffamily\scriptsize #2\end{minipage}}\par\vspace{0.5pt}}%
\newcommand{\frkey}[2]{{\setlength{\fboxsep}{1.2pt}\colorbox{fr#1Bg}{\frtt\vphantom{Ag}\frsym{#1}}}\kern1.2pt#2}%
\setlength{\parskip}{0pt}%
\par\hrule height 0.6pt\vspace{1.5pt}
\frhead{\textbf{Old}: first version, commit \texttt{25c8fe4} (Jan 2026)}{\textbf{New}: last of 4 versions, commit \texttt{e77d641} (Mar 2026)}
\frhunk{%
\frl{mod}{2}{0}{name: NoteOrganizer}}{%
\frl{mod}{2}{0}{name: note-{}organizer}}
\frhunk{%
\frl{del}{23}{0}{2. **Consult Rules**: Read the rule files located in \textasciigrave{}\frx{.cursor/rules/}\textasciigrave{} \frel}
\frl{del}{26}{3}{-{} \textasciigrave{}\frx{programming\_notes.mdc}\textasciigrave{}: For notes about coding, algorithms, etc.}}{%
\frl{mod}{20}{0}{2. **Load Project Memory**: \frel}
\frl{add}{21}{3}{-{} **Must read**: \textasciigrave{}\frx{knowledge/projects/\_index.md}\textasciigrave{} (project index)\frprobe}
\frl{mod}{24}{3}{\textgreater{} **Important**: Must use \frx{Read} tool \frel}}
\frhunk{%
\frl{mod}{39}{0}{7. **Update Changelog**: Add an entry to \textasciigrave{}\frx{knowledge/logs/changelog.md}\textasciigrave{} with timestamp, operation type, and description in Chinese.}}{%
\frl{mod}{47}{0}{9. **Update Changelog**: Add entry to \textasciigrave{}\frx{knowledge/logs/\_changelog.md}\textasciigrave{} following the changelog skill format.}}
\par\vspace{0.6pt}\noindent\mbox{\sffamily\scriptsize
\frkey{add}{added directive}\hspace{4.5pt}%
\frkey{del}{deleted directive}\hspace{4.5pt}%
\frkey{mod}{other: reworded, restructured, prose}\hspace{4.5pt}%
{\frtt\frx{\hspace{7pt}}}\kern1.2pt executable surface: path, command, tool, loading field}\par
\vspace{2pt}\hrule height 0.3pt\vspace{2pt}
\noindent{\scriptsize\renewcommand{\arraystretch}{0.94}\begin{tabular}{@{}>{\sffamily\bfseries}l@{\hspace{4pt}}>{\raggedright\arraybackslash}p{\dimexpr\columnwidth-30pt\relax}@{}}
Request & {\sffamily (generated; never mentions the directive)} \emph{I need to save a new note about the Q4 marketing strategy. Please categorize it into the correct project folder based on our current project structure.}\quad\textsf{\textbf{Check}} response must contain \texttt{knowledge/projects/\_index.md}\\
Result & {\sffamily compliant samples, none / old / new:} Qwen~3.5-9B 0/5, 0/5, 5/5;\quad all 16 models 0/80, 0/80, 43/80\\
\end{tabular}}\par\vspace{1.2pt}\hrule height 0.6pt
\Description{Side-by-side excerpt of the first and last version of a real SKILL.md file, with added, deleted and reworded lines marked, followed by the generated user request, the string the response must contain, and how many sampled answers comply under each condition.}
\caption{An illustrative revision of a real Skill (note-organizer in Kang-chen/kownledgeBase) and the probe built from its added directive on line 21. Grading is a literal string test.}
\label{fig:probe}\end{figure}

\paragraph{Probe types and population}
Besides the \npool{} added-directive probes (\naddedrepos{} repositories) we build \ndeleted{} \emph{deleted-directive} probes, with the two
versions in reverse order, and \ndilution{} \emph{dilution} probes around a directive line identical in both versions, which test whether
the rest of a revision changes compliance with a rule it leaves alone. Retained directives are less often prohibitions (\retnegin\% against \retnegout\%), conditional rules
(\retcondin\% against \retcondout\%) or permission guardrails than rejected ones (Table~\supretained). The Skills behind the probes are
revised more than the average Skill (median net growth \pdnetin{} against \pdnetall{} words, \pdcomin{} against \pdcomall{} revision
commits; Table~\supprobedist). Every effect is conditional on requests that need the rule.

\paragraph{Grader validation}\label{sec:validation}
A string check can pass a response that names the right identifier but uses it wrongly, and fail one that follows the rule in other words.
Two annotators, blind to model and condition, independently judged a stratified sample of \valn{} responses (Table~\supvalidation) on two
questions. Does the response follow the directive, and does it complete the task? Where both gave a yes or no, they agree at
$\kappa=\dfkappa$ on the first question (\dfdefn{} responses) and $\kappa=\tckappa$ on the second (\tcdefn). Against the directive-following
labels the grader has precision \dfpreclo--\dfprechi{} and recall \dfrecaone{} and \dfrecatwo{} (annotators 1 and 2), and on the 120
added-directive responses its recall falls from \dfrecold{} under the old version to \dfrecnew{} under the new one (annotator~1). Re-weighted
from this stratified sample to all main-set answers, both annotators' labels give a larger gain in directive following (about \dfeffa)
than the string check gives on the same answers (\dfmech). The adjudicator resolved the \tcadjn{} responses without an agreed task-correctness label
(\tcconf{} conflicting labels, \tcunc{} marked unclear by at least one annotator). This sample validates the Level-1 measure, and the primary
outcome evidence is the blind human judgement of agent artifacts in the sandbox (Table~\ref{tab:human}). Three software engineers (two with
master's-level and one with doctoral-level training) labelled this sample and the first sandbox round, two independently and one
adjudicating. Two further annotators, a machine-learning researcher and a doctoral-level researcher, labelled the second sandbox round.
No annotator or adjudicator in this study is an author of this paper or of the studied Skills. Claude Sonnet~4.5 (\texttt{claude-sonnet-4-5-20250929}, temperature 0) assigns
each deleted directive one of six deletion reasons. An annotator re-labelled 64 of them, and we report the hand-checked labels
(\S\ref{sec:deletion}).

\subsection{Model and Agent Settings}\label{sec:settings}
\emph{Models.} We select open-weight and proprietary models that cover several vendors, sizes and release dates. The \nmodels{} open-weight
instruction-tuned models span 3.8B to 32B parameters and six families (\famlist, and Gemma). All but one belong to six \emph{generation
ladders}. Each ladder holds successive releases of one family, so a comparison of generations never mixes in a change of vendor. This is
why we keep older releases such as Mistral 7B-v0.2. Gemma is evaluated only for breadth. Small open models let us run every model on every probe locally, and open
weights with fixed decoding settings make the runs reproducible. The five closed models are three Claude Sonnet releases (3.5, 3.7 and
4.5), Claude Haiku~4.5 and GPT-5.5, run on the 160-probe subset (\S\ref{sec:metrics}). In the sandbox we add Devstral-Small-2507, a 24B
open model from Mistral trained for coding agents, served with FP8 weights and cache. Other open models run in bf16. We disable thinking
(Qwen~3 and 3.5) and set GPT-5.5's reasoning effort to \emph{none}, so that every model answers under the same output budget. All runs use
temperature 0.7, with five samples per condition in single-step answers (three for closed models). In the sandbox we run three episodes
per cell, six for the Qwen models under on-demand loading and one for Sonnet~4.5. Table~\supmatrix{} shows which model ran in which
experiment.

\emph{Agent settings.} An agent is a model together with its harness~\cite{anthropic2026evals}, so we name the harness in every agent
setting. \emph{Single-step}: the prompt always names the Skill by its frontmatter name and description, as a harness lists installed
Skills, and the old and new conditions add the body, as a harness does when it activates the Skill~\cite{anthropic2026skillsdocs,agentskills2026client}.
Like~\cite{yang2026webdevskills}, we place only \texttt{SKILL.md} in the prompt, but the model answers once, without the retry of~\cite{yang2026webdevskills} or the tools of~\cite{liu2026wild}.
\emph{Multi-step}: a read-only tool agent lists, searches and reads the repository for up to eight turns, and an executing agent under the
published mini-SWE-agent scaffold~\cite{yang2024sweagent,miniswe2025} writes files and runs shell commands for up to twelve turns in a
network-less sandbox holding the repository at the new version's commit (\S\ref{sec:xbench}). We also let the agent decide whether to read
the Skill (\S\ref{sec:disclosure}), and pose single-step probes inside long contexts that resemble an agent's (\S\ref{sec:longctx}).

\subsection{Evaluation Metrics}\label{sec:metrics}
\emph{Effectiveness} has three levels. \emph{Directive compliance} in one answer is Level~1 (string check), the \emph{required action} of
an executing agent is Level~2, and \emph{final correctness} of the produced artifact is Level~3. Level~1 covers every model but also
credits an answer that repeats a token it was shown, so it includes \emph{information exposure} as well as \emph{directive following}. Let
$i$ index probes (in the sandbox, tasks) and $c\in\{\text{none},\text{old},\text{new}\}$ the Skill condition. Every condition gives the
Skill's name and description. Under \emph{none} the prompt has nothing more. Under \emph{old} and \emph{new} it also has the body of the old
or the new version. Let
$r=1,\dots,R$ index the repeated samples of one probe under one condition ($R=5$ in single-step answers, $R=3$ for closed models; in the
sandbox $r$ indexes episodes). $s_{icr}\in\{0,1\}$ is 1 when sample $r$ passes the check of the level being measured, and $N$ is the number
of probes a model answers. Then
\begin{equation}\label{eq:eff}
y_{ic}=\frac{1}{R}\sum_{r=1}^{R}s_{icr},\qquad E_c=\frac{1}{N}\sum_{i=1}^{N}y_{ic},\qquad \Delta E=E_{\text{new}}-E_{\text{old}},
\end{equation}
where $y_{ic}$ is the pass rate of probe $i$ under condition $c$, $E_c$ the model's mean pass rate under condition $c$, and $\Delta E$ the
effect of a revision, computed from the same probes under both versions. \emph{Tokens.} $I_{icr}$ and $O_{icr}$ are the input and output
tokens of sample $r$ (under the model's own tokenizer), and
\begin{equation}\label{eq:tok}
T^{\text{in}}_c=\frac{1}{N}\sum_{i=1}^{N}\frac{1}{R}\sum_{r=1}^{R}I_{icr},\qquad T^{\text{out}}_c=\frac{1}{N}\sum_{i=1}^{N}\frac{1}{R}\sum_{r=1}^{R}O_{icr},
\end{equation}
are the mean input and output tokens per response under condition $c$. $\Delta T^{\text{in}}=T^{\text{in}}_{\text{new}}-T^{\text{in}}_{\text{old}}$
and $\Delta T^{\text{out}}=T^{\text{out}}_{\text{new}}-T^{\text{out}}_{\text{old}}$ are the \emph{marginal cost} of a revision. The same
differences between old and none are the \emph{carrying cost} of a Skill body. We report the relative input change
$\Delta T^{\text{in}}/T^{\text{in}}_{\text{old}}$, the output change $\Delta T^{\text{out}}$, and
$\Delta T/\Delta E=(\Delta T^{\text{in}}+\Delta T^{\text{out}})/\Delta E$, the added tokens per additional compliant response. In agent
episodes tokens are summed over turns. \emph{Latency} is the mean wall-clock time per request, measured locally at batch size one.
\emph{Statistics.} Differences are paired over probes, with 95\% confidence intervals from a repository-clustered bootstrap (4{,}000
resamples), which resamples whole repositories so that probes from one repository stay together. Holm correction adjusts p-values for the
number of tests in a family. For claims of no difference we report a \emph{data-derived bound}, the larger absolute endpoint of the 90\%
interval. Differences larger than this bound are unlikely given the data, but the bound is not a margin fixed in advance.
\emph{Probe sets.} We built probes in two batches (317, then 79). All \nmodels{} open-weight models answer the same $N=\nmain$ probes, the
\emph{main set}. It holds the \nmainorig{} first-batch probes whose three conditions fit every open model's usable input (its context
window, capped at 8{,}192 tokens; 4{,}096 for OLMo-2) and \nmaindrawn{} drawn at random (seed 0) from the \nextfit{} second-batch probes
that also fit. The closed models, the latency runs and the long-context runs use a \emph{160-probe subset} drawn at random from the first
batch. The control experiments (\S\ref{sec:controls}) run on four models with 8{,}192-token windows and use the 292 first-batch probes
whose six prompt variants fit them. Without OLMo-2's limit, this set is larger than the 245 first-batch probes of the main set.

\section{RQ1: How Do Agent Skills Co-Evolve with Their Harness?}\label{sec:rq1}
\paragraph{Motivation and Approach}
A Skill reaches the model through a harness and points the agent to files, scripts and tools. If Skills co-evolve with their harness, a
Skill revision should change what the Skill asks the harness to read or run, and its commit should change harness configuration more often
than comparable commits. For the \nskills{} Skills we measure what each revision changes (\S\ref{sec:rq1rev}) and how the executable
references of a Skill change (\S\ref{sec:harness}). We then compare how often harness configuration changes in the same commit as a Skill
against commits matched by repository and commit-size band (\S\ref{sec:coevo}). We also read the documented motive and the authorship of
each revision commit.
\emph{Measures.} A rule-based classifier assigns each added or deleted line to one of ten content types by its Markdown form. The types
are code (fenced block), heading, table row, link or sub-document reference, numbered step, MUST/NEVER constraint, conditional bullet,
other bullet, formatting, and otherwise prose. Regular expressions over backticked spans extract the \emph{executable surface}, namely
repository paths, shell commands, agent tool names, frontmatter fields that control loading (e.g., \texttt{allowed-tools}) and ``before X,
read or run Y'' preconditions. On 200 hand-labelled added lines the classifier agrees with the annotator at $\kappa=\clfkappa$ (accuracy
\clfacc). It over-assigns code (precision \clfcodeprec) and misses most MUST/NEVER and conditional lines (recall \clfmustrec, \clfcondrec).
On the same lines plus 50 that it flags (ten per reference type), the extractor is precise for paths and commands (\expathprec,
\excmdprec) but sees only backticked spans (recall \expathrec, \excmdrec), so we report the references it detects. Tool-name counts are
unreliable (precision \extoolprec), so we do not report them. A file counts as harness configuration if it matches an agent-configuration
pattern (\texttt{CLAUDE.md}, \texttt{AGENTS.md}, \texttt{.claude/} and other agents' directories, Copilot instructions, MCP configuration
and plugin manifests; Table~\supharnessfiles). For every commit that touches a Skill we list the other files it changes and compare it with up to 20 random commits of
the same repository that touch no Skill. Skill commits change more files (median \coevmedskill{} against \coevmedbase), so we compare
within repository and within bands of files changed, weighting by Skill commits.

\subsection{Most revisions change a rule or procedure}\label{sec:rq1rev}
\begin{figure}[t]\centering
\begin{minipage}[t]{.495\linewidth}\centering\includegraphics[width=\linewidth]{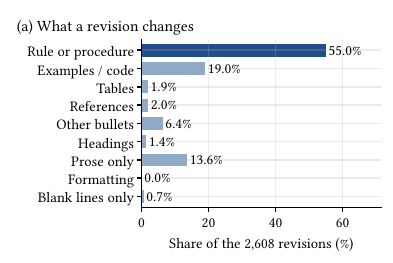}\end{minipage}\hfill
\begin{minipage}[t]{.495\linewidth}\centering\includegraphics[width=\linewidth]{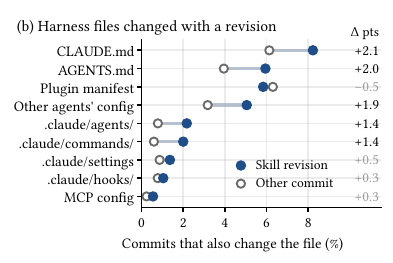}\end{minipage}
\Description{Two horizontal charts side by side. Left: share of revisions by the first type of change they contain, led by rule or
procedure changes at 55 percent, then examples or code and prose only. Right: for each kind of harness file, the share of Skill revision
commits and of comparable other commits that also change it, with the difference in percentage points.}
\caption{How Skills and their harness change. (a)~Each first/last pair counts once, under the first type in the order shown that it adds
or deletes a line of. Rule or procedure means a MUST/NEVER constraint, conditional rule or numbered step, as detected by the classifier
(\S\ref{sec:rq1}). (b)~Share of the \hfrevn{} Skill revision commits (filled) and of other commits of the same repository and size (open)
that also change each harness file; $\Delta$ in points, grey where the 95\% CI includes 0 (Table~\supharnessfiles).}\label{fig:growth}\end{figure}

Among the Skills that have a revision history (\nsampshare\% of the sampled files), revisions come often and in bursts (Table~\suprevisions). Consecutive versions are a median of under three days apart,
and \revlateshare\% of revision commits come more than a week after the first version. Although \hwordsup\% of revised Skills end longer,
the median net growth is \pdnetall{} words (Table~\supprobedist). Most revisions change a rule or procedure (Figure~\ref{fig:growth}a). Of the \revnpairs{} revisions,
\revrulepctx\% add or delete at least one MUST/NEVER constraint, conditional rule or numbered step,
\revcodepct\% change examples or code but no rule or procedure, and \revprosepct\% change only prose. By words, prose is the largest
addition, and code shrinks in net only because of a few large Skills (Figure~\supfiggrowth, Table~\supgrowth).

\paragraph{Documented motives and authorship}
The repository history rarely documents a model-related motive. Only \intentstrict\% of commit subjects name a model version, \intenthumanrate\%
of \intentn{} hand-labelled commits target a model or platform, and \mtpairs\% of pairs add or change a versioned model name or a
\texttt{model} field. These counts cover documented motives only (\S\ref{sec:threats}). Most commits add content (\intentnew\%), follow a
change in code or tooling (\intentsync\%), or correct an error (\intentcorr\%). An AI-agent or bot signature appears on \authshare\% of
revision commits, against \authnonskill\% of other commits in the same repositories and period.

\subsection{Few Skills gain executable references}\label{sec:harness}
Each executable reference type rises on average from first to last version, but from a small base (Table~\ref{tab:harness}), about
\abspathsd{} more paths and \abscmdsd{} more commands against \abswordsd{} more words per Skill. Most Skills leave their executable surface
unchanged. Skills get longer, but the share of paths and commands in them does not grow. The count per
1{,}000 words shows no detectable change (\densmain{} [\densmainlo, \densmainhi]). The same holds in the high-star set (\densflag{}
[\densflaglo, \densflaghi]) and with a higher-recall extractor that reads unquoted paths and commands (Table~\supextractor). The \texttt{user-invocable} field, which
turns a Skill into a command the user can call, rises from \uiallfirst\% to \uialllast\% of Skills, and doubles (\uiafterfirst\% to
\uiafterlast\%) among the \uiaftern{} Skills first written after the field appeared.

\begin{table}[t]\centering
\caption{Executable surface of a Skill in its first and last version (\npairs{} pairs). First, Last: per-Skill mean count; $\Delta$:
mean per-Skill change with a repository-clustered 95\% CI; Up / down: share of Skills whose count rose / fell. Counts are the references
the extractor detects in backticked spans.}\label{tab:harness}
\footnotesize
\ifdefined\rowcolor\def\tabharnessshade{\rowcolor[gray]{0.93}}\else\def\tabharnessshade{}\typeout{tab_harness: colortbl not loaded, group shading omitted}\fi
\begin{tabular}{lrrrr}\toprule
Count per Skill & First & Last & $\Delta$ [95\% CI] & Up / down\\\midrule
\tabharnessshade Repository paths referenced & 1.24 & 1.68 & +0.44 [+0.28, +0.62] & 18\% / 6\%\\
\tabharnessshade \quad of which scripts & 0.27 & 0.35 & +0.08 [+0.02, +0.15] & 6\% / 3\%\\
\tabharnessshade \quad of which reference documents & 0.69 & 0.94 & +0.25 [+0.13, +0.40] & 10\% / 3\%\\
Shell commands & 0.51 & 0.76 & +0.25 [+0.16, +0.35] & 12\% / 4\%\\
\tabharnessshade ``Before X, read/run Y'' steps & 0.05 & 0.07 & +0.02 [+0.01, +0.03] & 3\% / 1\%\\
\tabharnessshade Frontmatter control fields & 0.24 & 0.32 & +0.08 [+0.02, +0.14] & 8\% / 3\%\\
Words & 927.53 & 1028.62 & +101.09 [+46.11, +157.09] & 60\% / 24\%\\
\bottomrule
\end{tabular}
\end{table}

\subsection{Skill revisions touch harness files more often}\label{sec:coevo}
Of the \hfrevn{} commits that revise a Skill in the \hfnrepos{} repositories we could re-clone, \hfrevrate\% also change a
harness-configuration file, against \hfbaserate\% of other commits in the same repositories. Much of this gap reflects commit size.
Matched on repository and files changed, \hfmrate\% of revision commits and \hfmbase\% of other commits change harness configuration, a
difference of \hfmdiff{} points [\hfmlo, \hfmhi] (Figure~\ref{fig:growth}b). The excess comes from \texttt{CLAUDE.md} (\hfcmddiff),
\texttt{AGENTS.md} (\hfamddiff), the configuration of other agents such as Cursor and Codex (\hfothdiff), and Claude Code agent definitions
(\hfagtdiff) and command definitions (\hfcmdsdiff). Settings, hooks and MCP configuration change about as often as in other commits, and
plugin manifests no more often. Creating a Skill is coupled more tightly. Of the creation commits, \hfcrmrate\% change harness configuration,
against \hfcrmbase\% of matched commits (\hfcrmdiff{} [\hfcrmlo, \hfcrmhi]), across plugin manifests, settings and agent instructions
(Table~\supharnessfiles). A Skill commit also changes the Skill's own scripts and references \coevskillassetsmdiff{} points more often than a
matched commit (Table~\supcoevo). These are associations within commits, and Hereiz et al.~\cite{hereiz2026plugins} report a similar
coupling inside plugin Skills.

\insight{Skills change together with harness configuration. Revised Skills change often, mostly in their rules or procedures, and the
harness files around them change in the same commits more often than in comparable commits. Model-related motives are rarely documented.}

\section{RQ2: How Does Skill Evolution Affect Task Effectiveness?}\label{sec:rq2}
\paragraph{Motivation and Approach}
A maintainer revises a Skill to change an agent's behaviour, which can affect task effectiveness. We study rule changes, a narrower class within the \revrulepctx\% of revisions that RQ1's classifier marks as changing a rule or
procedure; that classifier misses most MUST/NEVER and conditional lines (recall \clfmustrec{} and \clfcondrec). Of the \npairs{}
revisions, 1{,}074 (41\%) add at least one directive that names a command, path or identifier. We focus on such rules because revisions
that change only prose or only examples barely move compliance with a Skill's existing rules. On the \texn{} dilution probes
whose revision changes only prose or examples, four models (Qwen~2.5-7B, Qwen~3-8B, Qwen~3.5-9B and Phi-4-mini) change compliance by
\texd{} [\texdlo, \texdhi], within an equivalence margin of 0.08 fixed before the run (per-model data-derived bounds at most
\texmaxbound), while the Skill body itself adds \texpos{} over none (Table~\suptexture). We measure
effectiveness at the three levels of \S\ref{sec:metrics} and supplement the string check with a token-only control, action traces and blind
judgements. Figure~\ref{fig:ladder} summarizes every setting of \S\ref{sec:single} (single answers) and \S\ref{sec:rq4}
(agents)\label{sec:ladder}. Effects are smaller in the agent settings we test than in single answers. On the same 36 tasks, final
correctness rises by less than the action rate, though not significantly (\S\ref{sec:human}).

\IfFileExists{fig_ladder.tex}{
\begin{figure}[t]\centering
\includegraphics[width=\columnwidth]{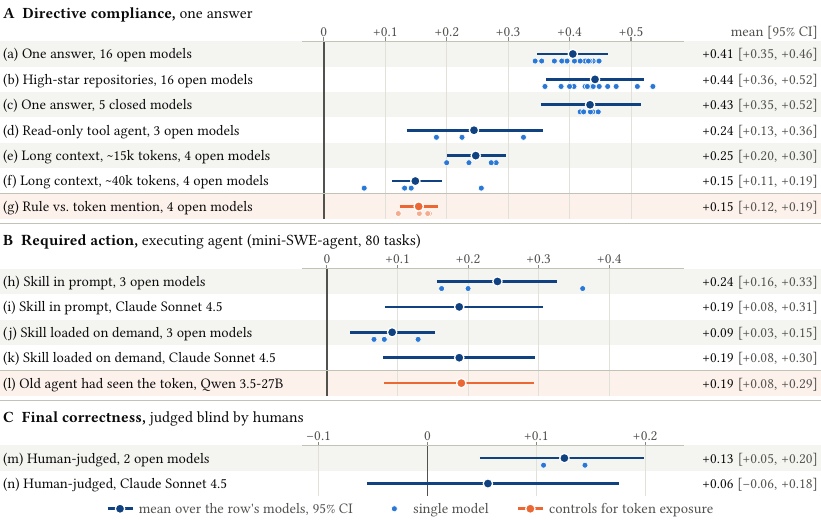}
\Description{Forest plot of new-minus-old differences with 95 percent confidence intervals in three panels: directive compliance in one answer, the required action of an executing agent, and human-judged final correctness. Each row is one setting, with model means and single models marked differently.}
\caption{New minus old in each setting. The three panels measure different constructs on their own axes and are not directly comparable; compliance~(A) includes information exposure. CIs: repository-clustered bootstrap. $n$: (a)~300 probes; (b)~65--87 per model (by context window); (c)~the 144 of 147 probes answered by all five; (d)~40; (e,~f)~149; (B)~80 tasks; (C)~151 and 72 judged pairs. Orange rows: (g)~the rule against a sentence naming only its token (292 probes); (l)~the 35 tasks where Qwen~3.5-27B's old agent had seen the token.}
\label{fig:ladder}
\end{figure}
}{}

\subsection{Single-Step Setting}\label{sec:single}
One model answers one request under each of three conditions: none, the old version and the new version (\S\ref{sec:settings}).

\subsubsection{Every model follows an added rule more often}
All \nmodels{} open-weight models follow the added rule more often, by a similar amount (Table~\ref{tab:effect}). We also ran the 147 probes of the 160-probe subset that fit every open model except the
4{,}096-token OLMo-2 (for GPT-5.5, the 144 it answered in both versions; Table~\supclosed). On them, per-model gains range from
\closedsonnetthreefivegain{} to \closedhaikufourfivegain{} for the five closed models and from \closedopenlo{} to \closedopenhi{} for the
open models. In the high-star set the per-model gains of the \flagmodels{} models range from \flaggainlo{} to \flaggainhi{} (65--87 probes
per model, by context window; Table~\supflagship). The gain does not differ detectably by the RQ1 characteristics of a revision, such as
its commit intent, a change in its executable references or a harness change in the same commits (smallest Holm-adjusted
$p=\linkminholm$; Table~\suptypology).

\begin{table}[t]\centering\caption{Compliance with the added rule under the old and new version, one answer, every model on the same $N=300$ probes.
$\Delta$: new $-$ old with repository-clustered 95\% CI (all $p<0.001$); the last two columns split $\Delta$ by whether the checked token is
new (193 probes) or already in the old version (107); $^{\circ}$: CI includes 0.}\label{tab:effect}\footnotesize
\begin{tabular}{llcrrrrrr}\toprule
& & & & & & & \multicolumn{2}{c}{$\Delta$ by token}\\\cmidrule(lr){8-9}
Family & Model & Released & Old & New & $\Delta$ & 95\% CI & new & present\\\midrule
Qwen & 2.5-7B & 2024-09 & 0.253 & 0.649 & +0.395 & [+0.335, +0.455] & +0.58 & +0.07\\
Qwen & 3-8B & 2025-04 & 0.280 & 0.719 & +0.439 & [+0.366, +0.510] & +0.66 & +0.04$^{\circ}$\\
Qwen & 3.5-9B & 2026-02 & 0.276 & 0.701 & +0.425 & [+0.361, +0.488] & +0.62 & +0.08\\
Qwen & 2.5-32B & 2024-09 & 0.270 & 0.703 & +0.433 & [+0.365, +0.502] & +0.62 & +0.10\\
Qwen & 3-32B & 2025-04 & 0.279 & 0.725 & +0.446 & [+0.379, +0.513] & +0.65 & +0.08\\
Qwen & 3.5-27B & 2026-02 & 0.279 & 0.710 & +0.431 & [+0.362, +0.500] & +0.61 & +0.10\\
\midrule
Phi & 3-mini & 2024-04 & 0.217 & 0.613 & +0.396 & [+0.334, +0.456] & +0.56 & +0.10\\
Phi & 3.5-mini & 2024-08 & 0.232 & 0.622 & +0.390 & [+0.335, +0.444] & +0.55 & +0.10\\
Phi & 4-mini & 2025-02 & 0.249 & 0.624 & +0.375 & [+0.310, +0.439] & +0.55 & +0.07$^{\circ}$\\
\midrule
Mistral & 7B-v0.2 & 2023-12 & 0.247 & 0.634 & +0.387 & [+0.327, +0.448] & +0.56 & +0.08\\
Mistral & 7B-v0.3 & 2024-05 & 0.245 & 0.641 & +0.396 & [+0.333, +0.460] & +0.59 & +0.05$^{\circ}$\\
\midrule
Llama & 3-8B & 2024-04 & 0.249 & 0.657 & +0.408 & [+0.343, +0.473] & +0.59 & +0.09\\
Llama & 3.1-8B & 2024-07 & 0.270 & 0.687 & +0.417 & [+0.347, +0.486] & +0.63 & +0.04$^{\circ}$\\
\midrule
OLMo & 2-7B & 2024-11 & 0.210 & 0.554 & +0.344 & [+0.288, +0.401] & +0.51 & +0.05$^{\circ}$\\
OLMo & 3-7B & 2025-11 & 0.230 & 0.585 & +0.355 & [+0.296, +0.416] & +0.51 & +0.08\\
\midrule
Gemma & 2-9B & 2024-06 & 0.242 & 0.690 & +0.448 & [+0.381, +0.517] & +0.63 & +0.11\\
\bottomrule\end{tabular}\end{table}

\subsubsection{Stating a rule beats naming its token}\label{sec:controls}
\begin{figure}[t]\centering
\includegraphics[width=\columnwidth]{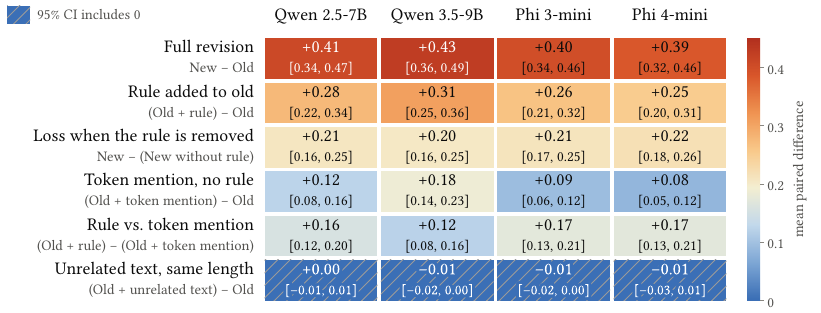}
\Description{Heat map with one column per control model and one row per contrast between prompt variants, such as the old version plus only the added rule against the old version. Each cell shows the mean paired difference in compliance and its confidence interval, coloured from cool blue for small differences to warm red for large ones; cells whose interval includes zero are hatched.}
\caption{Control conditions on the added-directive probes. Each row is A $-$ B as named under it (positive: A complies more). Cell: mean
paired difference in compliance with its repository-clustered 95\% CI. Colour runs from cool for small estimates to warm for large ones (0 to 0.45); hatched: CI includes 0. The first row re-estimates the full revision in this run. $n$: the 292 of 317 probes whose six prompt variants fit 8{,}192 tokens for each of the four control models (more than the 245 first-batch probes of the main set, which must also fit OLMo-2's 4{,}096 tokens).}
\label{tab:controls}\label{fig:controls}
\end{figure}

A first/last pair can differ in many lines, and an answer can pass the check by naming the token. Figure~\ref{fig:controls} separates these
with four extra versions of each Skill on the same probes. They are the old version plus only the added rule line, the new version without
it, the old version plus unrelated Skill text as long as the revision's growth, and the old version plus one sentence that names the token
without stating a rule.

On all four control models (Phi-3-mini, Phi-4-mini, Qwen~2.5-7B and Qwen~3.5-9B), the rule line carries most of the revision. Adding only
the rule line to the old Skill gives about two thirds of the full effect. Removing it from the new Skill lowers compliance by about
\ctlnodirmeanabs. On \unchn{} further probes built the same way around a directive line identical in both versions, removing that line
lowers compliance by a similar \unchlineabs{} (five open models).
Unrelated text of the same length has no detectable effect. Stating the rule adds \rvt{} [\rvtlo, \rvthi] beyond naming its token. On the
\rvtapresn{} probes whose token the old version already contains, it adds \rvtapres{} [\rvtapreslo, \rvtapreshi]. The conditions overlap,
because the rule line contains the token (\S\ref{sec:threats}).

\subsubsection{The gain comes from tokens new to the Skill}\label{sec:novelty}
For each probe we know whether its token already appeared in the old version (Table~\supnovelty). A new token lifts compliance by
\novnewlo{} to \novnewhi{} from an old-version baseline of \newtokold. An already-present one starts much higher (\presold) and lifts it by
\apresd{} [\apreslo, \apreshi] pooled over models (\apresn{} probes), so part of the difference may be \emph{headroom}, the room left above a
high old-version rate. In a random sample of 30 of the \apresn{} already-present cases, \nvrest{} restate a rule (e.g., moved from a table to a list) and
\nvexn{} add a rule about a token earlier shown only in an example. These \nvexn{} rules show no detectable gain (\nvex{} [\nvexlo, \nvexhi]; \nvexn{} rules $\times$ 16 models). A new
rule is also applied where it is not needed. On \mfkept{} requests for the same Skills that do not need the rule (written by the same
generator and screened by Qwen~3-32B), all six models we ran (Llama~3.1-8B, Mistral 7B-v0.3, Phi-4-mini, Qwen~3.5-9B and 27B, Gemma~2-9B)
use its token \mflo{} to \mfhi{} more often under the new version (Table~\supmisfire), against \mfrello{} to \mfrelhi{} where it is needed.

\subsubsection{Deleting a rule lowers compliance with it}\label{sec:deletion}
Removing a rule lowers compliance with it by 0.27--0.33 on each of eight models (six Qwen models, Phi-4-mini and OLMo-2;
Table~\supdel), about three quarters of what adding one does. By hand-checked reason (exploratory; Table~\supdelcat), deleting a duplicate
changes compliance by \ddup{} ($n=\ddupn$) and deleting a replaced rule by \drep{} ($n=\drepn$).

\subsubsection{Older and newer model generations gain similarly}\label{sec:generations}
A maintainer edits a Skill against the model they use, while users may run an older or smaller one. For every pair of generations within a
family we compute the interaction, the older model's gain minus the newer one's (\ninteractions{} comparisons; Table~\supequiv). No
generation difference is significant after Holm correction. The data-derived bound is \eqbound, about \eqfrac{} of the main effect
(Table~\supequivsens). It is \closedeqbound{} along the Sonnet ladder (Table~\supclosedladder) and \rvtbound{} for the directive-following
part. In a separate family of tests restricted to probes with a new token, one step differs (\ntstep, \ntint, Holm-adjusted
$p=\ntholm$), where the older model gains less.

\subsubsection{Gains shrink when the Skill fills the window}\label{sec:boundary}
The benefit of a new rule declines with \emph{Skill occupancy}, the ratio of Skill tokens to a model's usable context, as models degrade
with input length~\cite{levy2024same}. Across occupancy groups it falls from $\occlow$ at low occupancy to $\occhigh$ when the Skill fills
three quarters of the window (Table~\supocc). Below that, growth alone barely matters, and a growing document around an unchanged line has
no detectable effect (Table~\supdilution).

\subsubsection{Part of the gain survives in long contexts}\label{sec:longctx}
An agent's context also carries a system prompt, tool schemas and a history. We placed probes of the 160-probe subset inside such a
context, beyond the 8{,}192-token cap of short answers, and let the model answer once. The context holds a coding-agent system prompt with
fourteen tool schemas, the Skill where a harness would put it, and a prior exchange in which the agent explores the probe's repository,
followed by the request. Four open models
whose context windows hold them (Llama~3.1-8B, OLMo~3-7B, Phi-4-mini, Qwen~3.5-9B) answer the \lcprobes{} probes whose prompts fit all four
windows. In contexts of about \lcseslen k and \lclonglen k tokens the new version raises compliance by \lcgainrange{} (\lcsig{} of
\lcrows{} model--length cells significant), against \lcstrange{} in a short single answer on the same probes. About three fifths of the
short-answer effect survives at \lcseslen k tokens (95\% CI \lcretseslo--\lcretseshi\%) and about two fifths at \lclonglen k
(\lcretlonglo--\lcretlonghi\%). Where the context already shows the token but the old Skill lacks it (\lcinn{} probes), the revision still
adds \lcin{} [\lcinlo, \lcinhi].

\subsection{Multi-Step Setting}\label{sec:rq4}
We repeat the comparison with a read-only tool agent, an agent that executes commands in a sandbox, and the same agent loading the
Skill on demand. In each setting the Skill file is
removed from the checkout, so the only Skill the agent sees is the one we give it.

\subsubsection{Read-only agents follow the new rule more often}
On 40 probes drawn at random from the first batch (half of them prohibitions), agents that may list and read repository files for up to
eight turns comply more often with the new version. The gain is \agenticgainlo{} to \agenticgainhi{} on three open models (Qwen~2.5-7B,
Qwen~3.5-9B and Phi-4-mini; Table~\supagentic), significant for each.

\subsubsection{Sandbox agents take the required action more often}\label{sec:xbench}\label{sec:sota}
We built an executable benchmark of \xbntasks{} tasks whose added rule names a command, flag, file or identifier that exists in the
repository at the new version's commit (57 repositories; Table~\supsota), and the agent is asked to carry the request out. We accept at
three levels. \emph{Action} means that the required command was executed with its target present, the required file was read or created,
or a written file contains the required identifier. \emph{Execution} also requires that the command exits 0 and that written files parse.
\emph{Final correctness} is a per-task check, written for the \xbcn{} tasks where it could be, that the artifact has the content the
request asks for. The grader never reads the answer text, and every condition gives the Skill's name and description, so the conditions
differ only in the body. The protocol is the unmodified mini-SWE-agent scaffold~\cite{yang2024sweagent,miniswe2025} (its own prompt and
bash action format, twelve steps) in our network-less sandbox, with Qwen~3.5-9B and 27B, Devstral-Small-2507 and Claude Sonnet~4.5 on all
80 tasks.

\IfFileExists{tab_mswe.tex}{\begin{table}[t]\centering\caption{Sandbox under mini-SWE-agent~\cite{yang2024sweagent,miniswe2025}, three runs per cell (one for Sonnet~4.5). Action: rate under none (name and description only), old and new on all 80 tasks. Execution: the 28 tasks that are not read-only. Final: automated final correctness on the 36 tasks with a final check. Repository-clustered 95\% CIs.}\label{tab:mswe}\footnotesize
\setlength{\tabcolsep}{3.5pt}\begin{tabular}{lrrrr}\toprule
& Qwen~3.5-9B & Qwen~3.5-27B & Devstral Small & Claude Sonnet~4.5\\\midrule
Action: none / old / new & 0.32 / 0.42 / 0.58 & 0.40 / 0.42 / 0.62 & 0.34 / 0.44 / 0.80 & 0.49 / 0.62 / 0.81\\
\quad new $-$ old [95\% CI] & +0.16 [+0.06, +0.25] & +0.20 [+0.11, +0.29] & +0.36 [+0.24, +0.48] & +0.19 [+0.08, +0.31]\\
\midrule
Execution: old $\to$ new & 0.20$\to$0.33 & 0.19$\to$0.38 & 0.49$\to$0.74 & 0.64$\to$0.79\\
\quad $\Delta$ [95\% CI] & +0.13 [\ensuremath{-}0.04, +0.31] & +0.19 [+0.05, +0.36] & +0.25 [+0.08, +0.45] & +0.14 [\ensuremath{-}0.10, +0.40]\\
\midrule
Final: old $\to$ new & 0.19$\to$0.28 & 0.16$\to$0.31 & 0.39$\to$0.63 & 0.67$\to$0.69\\
\quad $\Delta$ [95\% CI] & +0.08 [\ensuremath{-}0.06, +0.24] & +0.15 [+0.02, +0.29] & +0.24 [+0.07, +0.44] & +0.03 [\ensuremath{-}0.18, +0.25]\\
\bottomrule\end{tabular}\end{table}}{}

With the new version every agent takes the required action more often (Table~\ref{tab:mswe}), by \faact{} [\faactlo, \faacthi] on average
over the four agents. Devstral, an open model trained for coding agents, gains most (\devA), and Sonnet~4.5 gains \sotamnewold{} from a higher
starting point (\sotamold{} under the old version).
\emph{Stale references.} The sandbox holds the repository at the new version's commit, so an old Skill can name files that a later code
change removed, as \texttt{changelog.md} became \texttt{\_changelog.md} in Figure~\ref{fig:probe}. Part of the gain may then be that of a
current Skill over a stale one, which is also what a maintainer who updates a Skill obtains. On the \stresn{} tasks where every path,
script, command and build target that the old Skill names still exists at the new commit, the action gain pooled over the four agents is
\stres{} [\streslo, \streshi] (per agent \stresmin{} to \stresmax), against \stall{} on all 80 tasks. It differs from that on the
\ststalen{} tasks with a missing reference by \stdiff{} [\stdifflo, \stdiffhi].
\emph{Information already available.} Every task's required token exists in the repository, and old-version agents see it in their own
command output in \rdseenlo--\rdseenhi\% of episodes on the \rdnotold{} tasks whose old version lacks it. Where Qwen~3.5-27B's old-version
agent saw the token, the new version still raises its action rate by \rdsawqq{} [\rdsawqqlo, \rdsawqqhi]. Where the old Skill already contained the token (\xbpresn{}
tasks), we detect no action gain (Qwen~3.5-9B: \xbpresq, 27B: \xbpresqq, Sonnet~4.5: \xbpresm).
\emph{Behaviour.}\label{sec:behaviour} With a Skill body, agents list directories less often, and with the new version they run fewer
commands (Table~\supbehaviour).

\subsubsection{Human-judged correctness rises for two open models}\label{sec:human}
\IfFileExists{tab_human.tex}{\begin{table}[t]\centering\caption{Human-judged final correctness, blind to model and version (one run per model, task and version). Left: the 36 tasks with a final check; right: all 80 tasks. Pairs with an artifact still unclear after adjudication are dropped: 3 of 72 in the first round (36 tasks, Qwen) and 14 of 168 in the second (44 tasks for Qwen, 80 for Sonnet~4.5). New$-$old with repository-clustered 95\% CIs.}\label{tab:human}\footnotesize
\resizebox{\linewidth}{!}{\begin{tabular}{lrrrrrr}\toprule
& \multicolumn{3}{c}{36 tasks (adjudicated)} & \multicolumn{3}{c}{All 80 tasks}\\\cmidrule(lr){2-4}\cmidrule(lr){5-7}
Model & Pairs & Old$\to$New & Human $\Delta$ [95\% CI] & Pairs & Old$\to$New & Human $\Delta$ [95\% CI]\\\midrule
Qwen~3.5-9B & 34 & 0.21$\to$0.35 & +0.15 [\ensuremath{-}0.03, +0.34] & 76 & 0.36$\to$0.50 & +0.14 [+0.03, +0.26]\\
Qwen~3.5-27B & 35 & 0.23$\to$0.34 & +0.11 [\ensuremath{-}0.03, +0.26] & 75 & 0.39$\to$0.49 & +0.11 [+0.01, +0.20]\\
Qwen, pooled & 69 & 0.22$\to$0.35 & +0.13 [+0.03, +0.26] & 151 & 0.37$\to$0.50 & +0.13 [+0.05, +0.20]\\
\midrule
Claude Sonnet~4.5 & -- & -- & -- & 72 & 0.60$\to$0.65 & +0.06 [\ensuremath{-}0.06, +0.18]\\
\bottomrule\end{tabular}}\end{table}}{}
Table~\ref{tab:human} is our primary outcome evidence. Annotators blind to model and version judged one pre-drawn run per model, task and
version in two rounds. Two annotators judged 36 tasks for the two Qwen models ($\kappa=\hpk$ where both gave yes or no), two further
annotators the other 44 tasks and all 80 for Sonnet~4.5 ($\kappa=\hthreek$), and an adjudicator, also blind, resolved unresolved artifacts
from their full content (annotators: \S\ref{sec:validation}). Over all 80 tasks, final correctness rises by \hsp{} [\hsplo, \hsphi] for the
two Qwen models, significantly for each (\hspleast{} to \hspmost{} however the still unclear artifacts are assigned). For Sonnet~4.5 it
rises by \hson{} [\hsonlo, \hsonhi], an estimate with too few pairs to be conclusive (\S\ref{sec:threats}). Pooled over the three judged
agents it rises by \fahum{} [\fahumlo, \fahumhi]. The automated final check, available for \xbcn{} tasks, also rises for the open models,
including Devstral (\devF{} [\devFlo, \devFhi]), which was not judged by humans, and does not move for Sonnet~4.5. On these \xbcn{} tasks
final correctness rises by less than the action rate, but the differences are not significant (\msglo{} to \msghi).

\subsubsection{On-demand loading keeps about half of the action gain}\label{sec:disclosure}
This setting lets the agent decide whether to read the Skill body; our sandbox shows only the first and last 2{,}000 characters of a read
(\S\ref{sec:uptake}). Real tools list a Skill by name and description and load its body only when the agent decides it needs
it. We re-ran the sandbox with only the version's name and description in the prompt and the body at its path in the checkout. Averaged
over the four agents, the action gain falls from \faact{} to \faod{} [\faodlo, \faodhi]. About \faret\% of it survives (95\% CI
\faretlo--\farethi\%). Sonnet~4.5's action gain is \discson{} under both loading modes. On demand its old Skill works worse (automated
final correctness \discsonFold{} against \xbcmsfinold{} preloaded) while the new one works about as well (\discsonFnew), so final
correctness rises by \discsonF{} [\discsonFlo, \discsonFhi] on the \discFtasks{} tasks with a check. The \ndiscopen{} open models keep about
\discpoolret\% of their gain (95\% CI \discretlo--\discrethi\%). Qwen~3.5-27B drops from \discfull{} to \discA, 9B from \sotaqnewold{} to
\discnineA, and Devstral from \devA{} to \devdisc. Sonnet~4.5 and Qwen~3.5-27B read the Skill at similar rates.

\subsubsection{Agents that read the rule do not always act on it}\label{sec:uptake}
The two agents open the Skill in \discreadq\% (Qwen~3.5-27B) and \discreads\% (Sonnet~4.5) of new-version episodes. Our sandbox shows the first and last 2{,}000 characters of a command's output, so a
read displays the added rule in only \discvisq\% and \discviss\% of reads. Once the rule was displayed, Sonnet
took the required action in \discutils\% [\discutilslo, \discutilshi] of episodes and Qwen in \discutilq\% [\discutilqlo,
\discutilqhi]. The task-paired difference is \discutilgap{} [\discutilgaplo, \discutilgaphi]. With the body in the prompt the two act on the rule in \bodyutils\% and \bodyutilq\%
of episodes, so the gap is not specific to on-demand loading. Qwen's on-demand drop thus does not come from lower uptake once the rule is seen.

\insight{An added rule changes behaviour mainly when it names a command or path that the old Skill did not mention, and agents act on such
a rule rather than only repeating its token. How much of the gain an agent keeps depends on whether and how it loads the Skill.}

\section{RQ3: How Does Skill Evolution Affect Cost?}\label{sec:rq3}
\paragraph{Motivation and Approach}
When the Skill body is in the context, every added line is read again on every turn, so a revision adds tokens even when it helps. We separate two costs. The \emph{marginal
cost} of a revision is what the new version costs beyond the old one. The \emph{carrying cost} of a Skill body is what the old version
costs beyond none, the name and description alone. We measure both in tokens (\S\ref{sec:metrics}), first for the rule changes that RQ2
studies, in single-step answers (\S\ref{sec:cost-single}) and multi-step agent episodes (\S\ref{sec:cost-multi}), and then for every type
of revision (\S\ref{sec:revcost}), and latency for single requests. For closed models we report token cost for the four Claude
models, whose reported usage we could match to our prompts.

\subsection{Single-Step Setting}\label{sec:cost-single}
\subsubsection{A revision adds input tokens but shortens the answer}
Input grows because the new version is longer (Table~\ref{tab:tokcost}), and
output \emph{shrinks}, because a model told which command to run answers more concisely than one that hedges. In \texttt{juxt/allium}, a
rule to run \texttt{allium check} shrinks the answer from 201 to 81 tokens and raises compliance from 0.07 to 0.99. The marginal cost of a revision is therefore almost entirely input. Counting input and output, a request
uses \dtokopenabs\% more tokens on the open models and \dtokclosedloabs{} to \dtokclosedhiabs\% more on the four Claude models
(Table~\ref{tab:costlevels}). At list prices, where output costs three to five times more than input, the bill rises less
(Table~\supdollars). Latency does not follow tokens. Measured locally at batch size one on three open models, the new version changes mean
latency by $\latpbpct\%$ to $\latqapct\%$ (Table~\suplatency), because decoding dominates and the answer gets shorter.

\begin{table}[t]\centering\caption{Mean tokens per request for the revisions of Table~\ref{tab:effect}, under each model's tokenizer. $\Delta T/\Delta E$: added input and output tokens per additional compliant response.}\label{tab:tokcost}\footnotesize
\begin{tabular}{llrrrrrrr}\toprule
& & \multicolumn{3}{c}{Input tokens} & \multicolumn{3}{c}{Output tokens} & \\\cmidrule(lr){3-5}\cmidrule(lr){6-8}
Family & Model & Old & New & $\Delta$ & Old & New & $\Delta$ & $\Delta T/\Delta E$\\\midrule
Qwen & 2.5-7B & 1,522 & 1,811 & +19\% & 160 & 149 & \ensuremath{-}11 & 703\\
Qwen & 3-8B & 1,522 & 1,811 & +19\% & 137 & 128 & \ensuremath{-}8 & 640\\
Qwen & 3.5-9B & 1,592 & 1,885 & +18\% & 173 & 159 & \ensuremath{-}15 & 655\\
Qwen & 2.5-32B & 1,522 & 1,811 & +19\% & 168 & 155 & \ensuremath{-}14 & 636\\
Qwen & 3-32B & 1,522 & 1,811 & +19\% & 126 & 120 & \ensuremath{-}7 & 633\\
Qwen & 3.5-27B & 1,592 & 1,885 & +18\% & 191 & 174 & \ensuremath{-}17 & 640\\
\midrule
Phi & 3-mini & 1,896 & 2,252 & +19\% & 141 & 122 & \ensuremath{-}19 & 850\\
Phi & 3.5-mini & 1,896 & 2,252 & +19\% & 206 & 191 & \ensuremath{-}15 & 873\\
Phi & 4-mini & 1,510 & 1,797 & +19\% & 136 & 125 & \ensuremath{-}12 & 733\\
\midrule
Mistral & 7B-v0.2 & 1,888 & 2,240 & +19\% & 207 & 201 & \ensuremath{-}6 & 891\\
Mistral & 7B-v0.3 & 1,888 & 2,240 & +19\% & 195 & 188 & \ensuremath{-}7 & 871\\
\midrule
Llama & 3-8B & 1,509 & 1,798 & +19\% & 162 & 155 & \ensuremath{-}7 & 692\\
Llama & 3.1-8B & 1,509 & 1,798 & +19\% & 160 & 151 & \ensuremath{-}9 & 670\\
\midrule
OLMo & 2-7B & 1,509 & 1,802 & +19\% & 219 & 217 & \ensuremath{-}2 & 844\\
OLMo & 3-7B & 1,509 & 1,802 & +19\% & 213 & 204 & \ensuremath{-}9 & 800\\
\midrule
Gemma & 2-9B & 1,642 & 1,945 & +18\% & 114 & 103 & \ensuremath{-}11 & 652\\
\bottomrule\end{tabular}\end{table}

\subsubsection{One rule--model pair in five adds tokens with no detected gain}
For each added rule and each model, a rule--model pair, we checked whether compliance changed and whether tokens increased. With five
samples per condition, compliance moves in steps of 0.2. Compliance improves in \helpedpct\% of pairs. In \overheadpct\%, the prompt grows
but we detect no change in compliance, and we call these \emph{overhead candidates} (repository-clustered 95\% CI \ovcilo--\ovcihi\%;
\ovthrlo--\ovthrhi\% under six alternative thresholds). In \flatpct\% neither compliance nor cost changes, and in \downpct\% compliance
falls. Of the \ovdirn{} rules that at least half the models classify as overhead, \ovred{} are \emph{already followed}, meaning that the
models did what the rule says without it (mean old compliance at least 0.5). The other \ovineff{} are \emph{ineffective}, ignored whether
present or not. The first are candidates for review and warrant controlled removal tests before deletion; the second call for rewriting.

\subsection{Multi-Step Setting}\label{sec:cost-multi}\label{sec:dollars}
\begin{table}[t]\centering\caption{Marginal cost of a revision (new $-$ old) and carrying cost of a Skill body (old $-$ none) in a single answer, a long-context single answer (\S\ref{sec:longctx}) and a sandbox episode under mini-SWE-agent, where the body is re-sent every turn (Qwen~3.5-9B and 27B, Devstral and Sonnet~4.5). Latency: local, batch size one. $^\dagger$Every 95\% CI spans 0.}\label{tab:costlevels}\footnotesize
\begin{tabular}{llrr}\toprule
Setting & Unit & Revision & Skill body\\\midrule
Single answer, 16 open models & input + output tokens & +17\% & --\\
Single answer, 4 Claude models & input + output tokens & +24 to +27\% & --\\
Single answer, 3 open models & latency & \ensuremath{-}7.2 to \ensuremath{-}1.6\% & --\\
Long context (15k--40k), 4 models & input tokens & +1.1 to +3.5\% & --\\
Sandbox episode, 4 agents & tokens per episode & \ensuremath{-}2.3 to +3.0\%$^\dagger$ & +37 to +57\%\\
\bottomrule\end{tabular}\end{table}

\subsubsection{A revision adds no detectable cost per episode}
Over the 80 sandbox tasks, the new version changes the tokens of an episode by \agcostlo{} to \agcosthi\% across the four agents
(\famarg\% on average). Every interval spans zero and lies within \agcostcilo{} to \agcostcihi\%, so an increase of more than about
\agcostcihi\% is excluded. The longer prompt is offset by fewer steps (\S\ref{sec:behaviour}).

\subsubsection{Carrying a Skill body costs far more than revising it}
The old version costs \agskillloabs{} to \agskillhiabs\% more tokens per episode than none (\facarry\% on
average over the four agents). The body is re-sent on every turn, and agents with a Skill body take fewer turns (\S\ref{sec:behaviour}),
but their episodes still use more tokens (per-episode list-price costs: Table~\supsota).

\subsection{Every revision type adds input, prose-only the least}\label{sec:revcost}
Counting the Skill text our prompts insert with each open model's tokenizer over all \npairs{} revisions (input only, no model run;
Table~\suprevcost), a revision that changes a rule or procedure adds a median of \rcrulemed{} tokens and one that changes only prose
\rcprosemed. On the \texn{} dilution probes of RQ2, revisions that change only prose or examples add \texin{} input and \texout{} output
tokens per request.

\insight{The cost of a revision is mostly added input. In a single answer it adds \inplo--\inphi\% input tokens and shortens the answer
only slightly; in an agent episode fewer steps leave no detectable change. Loading the Skill body costs far more. For one rule--model pair
in five we detect no gain in compliance, only the added tokens.}

\section{Discussion}
\subsection{Threats to Validity}\label{sec:threats}
\paragraph{Construct validity}
Level-1 compliance checks the rule's own token, so it includes information exposure and credits an answer that repeats a token it was
shown. The control conditions separate exposure from directive following (\S\ref{sec:controls}), but the rule line contains the token, so
the conditions overlap and do not split the effect exactly. The same holds for the split by whether the old Skill named the token
(\S\ref{sec:novelty}). In the validation sample the check misses more compliant answers under the new version, and annotators saw a larger
gain than the check (\S\ref{sec:validation}). Table~\supannot{} adds sensitivity analyses, including one with a further annotator's judgements. At five
samples per condition, equal old and new compliance is compatible with a small real change, so an overhead candidate is not shown to have
no effect. Subsets defined by old-version behaviour (\S\ref{sec:xbench}) are descriptive. The automated final check inspects artifacts
rather than running tests and agrees with blind annotators only moderately ($\kappa=\hvkflo$--\hvkfhi), so the outcome claim rests on
blind human judgement.
\paragraph{Internal validity}
Under right truncation, which cuts the end of an over-long prompt, a prompt longer than a model's usable input loses the user request
(\truncover{} of \truncn{} probe--model prompts). We therefore score all \nmodels{} open-weight models on the same \nmain{} probes with the
request intact, and using all 304 eligible probes changes the main result only from \mainthreeoh{} to \mainthreeofour. Fork duplication does
not inflate the counts (six first and no last versions are byte-identical across repositories). Other ways of delivering the body change
the single-step gain by a few points at most, and an in-house sandbox harness gives gains of the same sign (Table~\supxbenchcommon). Probes are generated by Qwen~3.5-27B, and Qwen models gain
slightly more on them than the other \famon{} models (\famq{} against \famo, difference \famd{} [\famdlo, \famdhi]). On \phixn{} probes whose
requests Phi-4 wrote, the \phixnon{} non-Qwen models gain \phixlo{} to \phixhi, and three Qwen models gain \phixedge{} more than the others
(\phixedgemain{} on the main set, a change of \phixdd{} [\phixddlo, \phixddhi]; Table~\supgenerator).
\paragraph{Maintainer intent}
Maintainers may revise a Skill because of a new model or harness without saying so, so our RQ1 counts cover only documented motives. RQ2
tests whether revisions work across model generations rather than why they were made.
\paragraph{External validity}
Our findings apply to concrete, mechanically checkable rules. The literal filter keeps \nkeeppct\% of candidate directives and
under-represents prohibitions, conditional rules and permission guardrails (Table~\supretained), which a string check cannot grade. Open
models span 3.8B--32B parameters, the closed models come from two vendors, and models answered without reasoning. Reasoning models
may follow rules differently. The sandbox uses one published scaffold rather than production agents, and it shows only
the first and last 2{,}000 characters of a command's output, so on-demand gains for long Skills are likely lower than under harnesses that
display whole files. The corpus skews to moderately popular repositories (none above \corpusmaxstars{} stars in the main corpus), and the
high-star set (\S\ref{sec:corpus}) covers \flagrepos{} repositories above 1{,}000 stars with fewer probes. Most ``old'' versions post-date
the older models, so authorship time and generation are not fully crossed.
\paragraph{Statistical conclusion validity}
The generation result is a negative finding. Its data-derived bound is about \eqfrac{} of the main effect, and smaller interactions remain
possible. The Sonnet~4.5 outcome estimate is underpowered. With \sonpairs{} pairs, the power to detect an effect as large as the Qwen effect
(+0.13) is about \sonpower{} (minimum detectable effect \sonmde), and about \sonpairsneed{} pairs would give 80\% power.

\subsection{Implications for Skill Maintainers}\label{sec:need}
Every effect in RQ2 is conditional on a request that needs the rule. To estimate how often real work does, we sampled \needllmpairs{}
in-scope commit--rule pairs, up to 30 later commits per added rule that touch the Skill's files or the paths it names (120 rules, 74
repositories). An annotator labelled 100 of these pairs by hand, drawn stratified by the judgement of
Qwen~3-32B (greedy decoding) on whether completing the commit needs the rule, and reweighted them to all \needllmpairs{} pairs. Of these
commits, \hneed\% need the rule (95\% CI \needcilo--\needcihi\%; \hneedlo--\hneedhi\% by how unclear pairs are counted; Table~\supneedci).
Automated proxies under-count (Table~\supneed). As a scenario, suppose that the conditional effects hold unchanged on in-scope requests that
need the rule and are zero on those that do not. An in-scope request would then gain about \hneedgain{} in single-answer compliance
(0.\hneed{} $\times$ \maineffect; 95\% CI of the product \needgainlo--\needgainhi) and about \stneedgain{} in the required action of the four
sandbox agents (0.\hneed{} $\times$ \stallgain), while every request reads the revision's tokens. Three practices follow for maintainers.
Add a rule when it gives the agent information the Skill lacks. Before deleting a rule, check that the agent should no longer follow it,
since deletions are followed too. Size a Skill for the smallest model that will read it, since the model's release year matters less.

\subsection{Relation to Prior Findings}
Jiang et al.~\cite{jiang2026demystifying} find that Skills act mostly as procedural anchors rather than as knowledge injection. For a single
revision, we find the gain concentrated in rules whose token the old Skill did not mention, which we read as new information. Dong
et al.~\cite{dong2026harmful} find that Skills induce excessive verification, whereas our agents explore less.

\section{Conclusion and Future Work}
Revised Skills change often, mostly in their rules, and harness configuration changes with them. Added rules change what agents do,
mainly when they name a command or path the old Skill did not mention, with similar gains across model generations and smaller gains
under on-demand loading. In the sandbox, agents take the required action more often, and blind judges see modestly higher final
correctness. Stating a rule adds compliance beyond naming its token, unrelated text of the same length has no detectable effect,
and deleting a rule lowers compliance by about three quarters of what adding one gains. A revision adds mostly input tokens and no
detectable cost per episode, while loading the Skill body costs far more.

We see five directions for future work. First, graders that judge agent behaviour, such as whether a forbidden command ran, could
grade the prohibitions, conditional rules and permission guardrails that a string check cannot. Second, natural task streams would
give the net effect of a revision, since about \hneed\% of later in-scope commits need the added rule (\S\ref{sec:need}).
Third, production harnesses and reasoning models may load and follow Skills differently from those we tested. Fourth, a
pre-merge check could run a revision's probes and flag rules that add tokens with no detected gain. Fifth, Skills first written after
the newer models would separate authorship time from model generation, which our corpus does not fully cross. Each direction can start from the probe and task manifests, per-sample results and
blind judgements in our replication package (\S\ref{sec:data}).

\clearpage
\section{Data Availability}\label{sec:data}
Replication package: \url{https://figshare.com/s/1fd1668be2b3b5081ad6}, with the supplementary tables (\texttt{supplement.pdf}),
probe and task manifests, per-sample results and the scripts behind every table and figure.

\bibliographystyle{ACM-Reference-Format}
\bibliography{refs}

\end{document}